\documentclass{pasj00}
\draft
\usepackage{pifont}
\usepackage{multirow}
\usepackage{lscape}
\usepackage{ccaption}

\begin{document}

\SetRunningHead{K. Torii et al.}{Temperature and Density in the Foot Points of the Molecular Loops in the Galactic Center}
\Received{2010/3/12}
\Accepted{2010/03/17}

\title{Temperature and Density in the Foot Points of the Molecular Loops in the Galactic Center; Analysis of Multi-J Transitions of \atom{C}{}{12}\atom{O}{}{}($J$=1--0, 3--2, 4--3, 7--6), \atom{C}{}{13}\atom{O}{}{}($J$=1--0) and \atom{C}{}{}\atom{O}{}{18}($J$=1--0)}

\author{Kazufumi \textsc{Torii},\altaffilmark{1} 
        Natsuko \textsc{Kudo},\altaffilmark{1}
        Motosuji \textsc{Fujishita},\altaffilmark{1}
        Tokuichi \textsc{Kawase},\altaffilmark{1} \\
        Takeshi \textsc{Okuda},\altaffilmark{1}
        Hiroaki \textsc{Yamamoto},\altaffilmark{1}
        Akiko \textsc{Kawamura},\altaffilmark{1}
        Norikazu \textsc{Mizuno},\altaffilmark{2}\\
        Toshikazu \textsc{Onishi},\altaffilmark{3} 
        Mami \textsc{Machida},\altaffilmark{1}
        Kunio \textsc{Takahashi},\altaffilmark{4}
        Satoshi \textsc{Nozawa},\altaffilmark{5} \\
        Ryoji \textsc{Matsumoto},\altaffilmark{6}
        J\"urgen \textsc{Ott},\altaffilmark{7}
        Kunihiko  \textsc{Tanaka},\altaffilmark{8}
        Nobuyuki  \textsc{Yamaguchi},\altaffilmark{9}\\
        Hajime  \textsc{Ezawa},\altaffilmark{10}
        J\"urgen \textsc{Stutzki},\altaffilmark{11} 
        Frank \textsc{Bertoldi},\altaffilmark{12}\\ 
        Bon-Chul \textsc{Koo},\altaffilmark{13}
        Leonardo \textsc{Bronfman},\altaffilmark{14}
        Michael \textsc{Burton},\altaffilmark{15}
        Arnold O. \textsc{Benz},\altaffilmark{16}\\
        Hideo \textsc{Ogawa},\altaffilmark{3}
        and 
        Yasuo \textsc{Fukui}\altaffilmark{1}
        }

\altaffiltext{1}{Department of Astrophysics, Nagoya University, Furo-cho, Chikusa-ku, Nagoya, Aichi 464-8602}
\altaffiltext{2}{National Astronomical Observatory of Japan, Osawa, Mitaka, Tokyo 181-8588}
\altaffiltext{3}{Department of Physical Science, Osaka prefecture University, Sakai, Osaka 599-8531}
\altaffiltext{4}{Japan Agency for Marine-Earth Science and Technology, Kanazawa-ku, Yokohama, Kanagawa 236-0001, Japan}
\altaffiltext{5}{Department of Science, Ibaraki University, 2-1-1 Bunkyo, Mito, Ibaraki 310-8512}
\altaffiltext{6}{Faculty of Science, Chiba University, Inage-ku, Chiba 263-8522}
\altaffiltext{7}{National Radio Astronomy Observatory, P.O. Box O, 1003 Lopezville Road, Socorro, NM 87801, USA}
\altaffiltext{8}{Institute of Science and Technology, Keio University, 4-14-1 Hiyoshi, Yokohama, Kanagawa 223-8522}
\altaffiltext{9}{Open Technologies Research Laboratory, 6-1-21 Hon-komagame, Bunkyo, Tokyo 113-0021}
\altaffiltext{10}{Nobeyama Radio Observatory, National Astronomical Observatory of Japan, Minamimaki, Minamisaku, Nagano 384-1305}
\altaffiltext{11}{KOSMA, I. Physikalisches Institut, Universit\"at zu K\"oln, Z\"ulpicher Stra$\beta e$ 77, 50937 K\"oln, Germany}
\altaffiltext{12}{Radioastronomisches Institut der Universit\"at Bonn, Auf dem H\"ugel 71, 53121 Bonn, Germany}
\altaffiltext{13}{Department of Physics and Astronomy, Seoul National University, Seoul 151-742, Korea}
\altaffiltext{14}{Departmento de Astronom\'ia, Universidad de Chile, Casilla 36-D, Santiago, Chile}
\altaffiltext{15}{School of Physics, University of New South Wales, Sydney 2052, NSW, Australia}
\altaffiltext{16}{Institute of Astronomy, ETH Zurich, 8093 Zurich, Switzerland}

\email{torii@a.phys.nagoya-u.ac.jp, fukui@a.phys.nagoya-u.ac.jp}


\maketitle

\begin{abstract}
\citet{fuk2006} discovered two molecular loops in the Galactic center and argued that the foot points of the molecular loops, two bright spots at both loops ends, represent the gas accumulated by the falling motion along the loops, subsequent to magnetic flotation by the Parker instability. We have carried out sensitive CO observations of the foot points toward $l$=356$^\circ$ at a few pc resolution in the six rotational transitions of CO; \atom{C}{}{12}\atom{O}{}{}($J$=1--0, 3--2, 4--3, 7--6), \atom{C}{}{13}\atom{O}{}{}($J$=1--0) and \atom{C}{}{}\atom{O}{}{18}($J$=1--0). The high resolution image of \atom{C}{}{12}\atom{O}{}{} ($J$=3--2) has revealed the detailed distribution of the high excitation gas including U shapes, the outer boundary of which shows sharp intensity jumps accompanying strong velocity gradients. An analysis of the multi-J CO transitions shows that the temperature is in a range from 30--100 K and density is around $10^3$--$10^4$ cm$^{-3}$, confirming that the foot points have high temperature and density although there is no prominent radiative heating source such as high mass stars in or around the loops. We argue that the high temperature is likely due to the shock heating under C-shock condition caused by the magnetic flotation. We made a comparison of the gas distribution with theoretical numerical simulations and note that the U shape is consistent with numerical simulations. We also find that the region of highest temperature of $\sim100$ K or higher inside the U shape corresponds to the spur having an upward flow, additionally heated up either by magnetic reconnection or bouncing in the interaction with the narrow neck at the bottom of the U shape. We note these new findings further reinforce the magnetic floatation interpretation.
\end{abstract}

\section{Introduction}
The central molecular zone (hereafter CMZ, \cite{mor1996}) is located in the inner 300 pc of the Galactic center and contains the Sgr A and Sgr B2 molecular clouds, two outstanding features in the CMZ (e.g., \cite{sco1975}; \cite{fuk1977}; \cite{gus1983}). The molecular gas in the CMZ is characterized by high kinetic temperature from 30 K to 300 K (e.g., \cite{rod2001}; \cite{hut1993}; \cite{mar2004}; \cite{oka2005}; \cite{nag2007}), and high density around 10$^4$ cm$^{-3}$ (e.g., \cite{tsu1999}). The molecular gas in the CMZ also shows violent motions with an rms velocity dispersion of 15--30 km s$^{-1}$, much larger than those of the typical molecular clouds of few km s$^{-1}$ outside the central kpc (\cite{mor1996}; \cite{gus2004}). It has been suggested that supernova explosions may be responsible for these properties, but the total star formation efficiency is too small to account for the required abundance of OB stars \citep{mor1996}. The origin of these two peculiar properties, high temperatures and violent motions, has been puzzling since its discovery in the 1970's. It is important to understand the physical properties of the molecular gas to understand star formation in the Galactic center and, consequently, the evolution of the Galaxy. 

There are several molecular features with very broad velocity widths outside the CMZ including Clumps 1 and 2 \citep{ban1977} and the $l=5.5^\circ$ cloud \citep{bit1997}. These clouds are located outside the CMZ, perhaps distributed in the central 1 kpc, but they have not yet been given as much attention as in the CMZ.

\begin{figure}
  \begin{center}
    \FigureFile(75.31mm,161.09mm){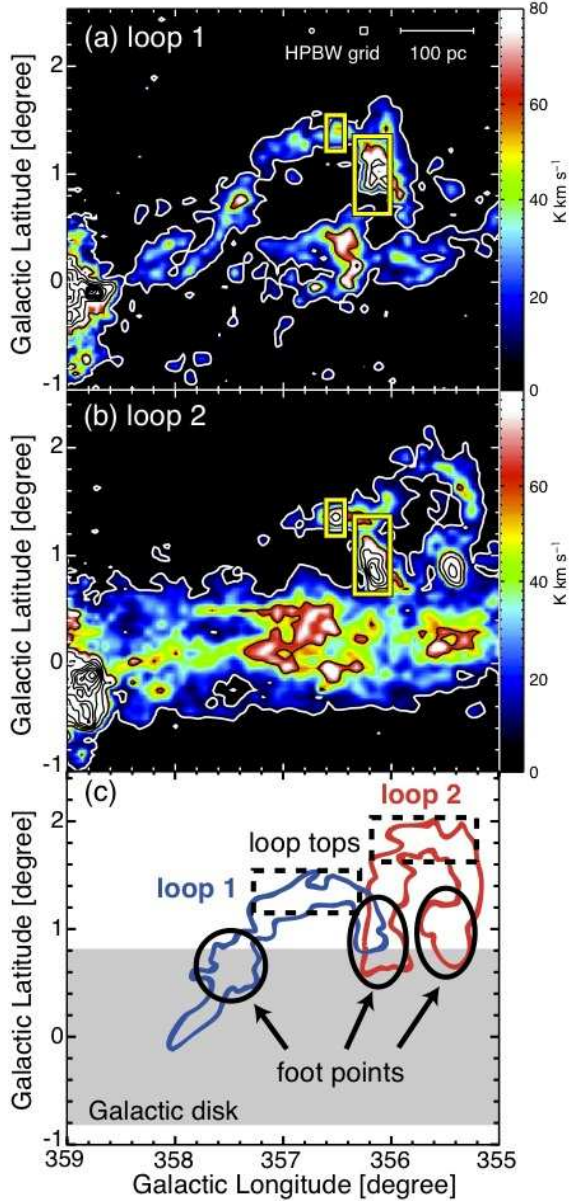}
  \end{center}
  \caption{(a--b) Integrated intensity distributions of loops 1 and 2 in $^{12}$CO($J$=1--0) obtained by NANTEN 4m telescope \citep{fuk2006}. Yellow boxes in each image show the observed region in $^{12}$CO($J$=3--2) by ASTE. (a) Loop 1 : The integration range in velocity is from $-180$ to $-90$ km s$^{-1}$. Contours are illustrated from 7 K km s$^{-1}$ (white) with an interval of 50 K km s$^{-1}$. (b) Loop 2 : The integration range in velocity is from $-90$ to $-40$ km s$^{-1}$. Contour levels are the same as that in (a). (c) Schematic image of loops 1 and 2. The foot points and tops of the loops are depicted by circles and dashed boxes.}\label{fig:loop12.ii+sch}
\end{figure}

Recently, Fukui et al. (2006; hereafter F06) discovered loop-like molecular features, hereafter loops 1 and 2, toward $l$=355$^\circ$--359$^\circ$ outside the CMZ (Figure \ref{fig:loop12.ii+sch}) and showed that the two loops each have two foot points on the both ends; the foot points are bright molecular condensations and show velocity widths as large as $\sim50$ km s$^{-1}$. F06 suggested that the two loops are formed by magnetic flotation driven by the Parker instability and showed that numerical magneto-hydrodynamics (MHD) calculations reproduce the loops successfully. This discovery is the first observational verification of the Parker's instability in the Galactic scale 40 years after the prediction by \citet{par1966}. In the scenario of F06, the two foot points are interpreted as the accumulated gas formed by the down-falling motion by the stellar gravity. Details of such a signature has already been shown by numerical simulations by \citet{mat1988}. F06 argues that the relatively strong magnetic field of 150 $\mu$G, which is eventually a result of the strong gravitation in the inner 1 kpc of the Galaxy, makes it possible to create the loops throughout the Galactic center. 

Subsequently, \citet{tor2009} made a detailed analysis of the NANTEN \atom{C}{}{12}\atom{O}{}{}($J$=1--0) and \atom{C}{}{13}\atom{O}{}{}($J$=1--0) datasets and revealed further details including helical distributions in loops 1 and 2 as well as associated \atom{H}{}{}\emissiontype{I} and dust features. \citet{fuj2009} presented the discovery of loop 3, another magnetically floated loop, in the same direction with loops 1 and 2 in a positive velocity range. \citet{mac2009} presented a three-dimensional global MHD simulations of the central 2-kpc magnetized gas disk, and \citet{tak2009} detailed two-dimensional local MHD simulations of the magnetic loops. These follow-up works offer further supports for the magnetic flotation picture.

In the magnetic flotation picture, the floated gas falls back down to the galactic plane along the loop driven by stellar gravity, and then collides with the nuclear gas disk. If the falling speed exceeds the sound speed, shocks must occur, leading to violent gas motions with enhanced density and kinetic temperature caused by compression and heating by the shocks. This is a conversion of the magnetic energy into the kinetic and thermal energy of the gas, and the magnetic flotation may provide an interpretation of physical conditions and kinematics of the central molecular gas. In order to test this scenario, it is important to reveal the distribution of temperature and density of the loop, in particular, in the foot points. 

In this paper, we present and discuss the results of pc-scale observations in six rotational transitions of \atom{C}{}{12}\atom{O}{}{} ($J$=1--0, 3--2, 4--3, 7--6),  \atom{C}{}{13}\atom{O}{}{}($J$=1--0) and  \atom{C}{}{}\atom{O}{}{18}($J$=1--0) toward two of the foot points with the ASTE, Mopra, NANTEN2 mm/sub-mm telescopes. We adopt a distance of 8.5 kpc to the Galactic center. Details of the observations are given in section 2 and the results are given in section 3. An analysis of the line radiative transfer is given in section 4, a discussion is in section 5, and the conclusions are presented in section 6.

\section{Observations}
We observed \atom{C}{}{12}\atom{O}{}{}($J$=1--0), \atom{C}{}{13}\atom{O}{}{}($J$=1--0) and \atom{C}{}{}\atom{O}{}{18}($J$=1--0) emission lines with the Mopra 22 m telescope and the \atom{C}{}{12}\atom{O}{}{}($J$=3--2) line with the ASTE 10m telescope. In addition we observed the \atom{C}{}{12}\atom{O}{}{}($J$=4--3, 7--6) lines with the NANTEN2 4m telescope. The \atom{C}{}{12}\atom{O}{}{}($J$=3--2) observing areas were selected using the NANTEN \atom{C}{}{12}\atom{O}{}{}($J$=1--0) dataset (Figure \ref{fig:loop12.ii+sch}). We produced a complete view of the foot points of the loops in \atom{C}{}{12}\atom{O}{}{}($J$=3--2) at a 40$''$ grid with a 22$''$ beam size. The Mopra observations were limited to a subset of the area obtained with ASTE, but covered the brightest features associated with the foot points. NANTEN2 observations were carried out toward four small regions centered on $J$=3--2 features identified in the ASTE results (Figure \ref{fig:lb.3-2.col}). The details of these observations are as follows (see also Tables \ref{table:obsline}, \ref{tab:obspsw} and \ref{tab:obsotf}).

\subsection{Mopra observations}
 Observations of \atom{C}{}{12}\atom{O}{}{}($J$=1--0), \atom{C}{}{13}\atom{O}{}{}($J$=1--0), and \atom{C}{}{}\atom{O}{}{18}($J$=1--0) were carried out by using the 22m ATNF Mopra mm telescope in Australia, during September 2007 and August 2008. On-the-fly (OTF) mode was used, with a unit field of 4$'$ $\times$ 4$'$ and we scanned in both longitudinal and latitudinal directions separately to minimize scanning effects. The telescope had a half-power beamwidth (HPBW) of 33$''$ at 100 GHz, which corresponds to 1.4 pc. The typical system noise temperature, $T_{\mathrm{sys}}$, was 500 K in the single side band (SSB). The Mopra telescope was also equipped with backend system "MOPS", providing 4096 channels across 137.5 MHz in each of the two orthogonal polarizations. The effective velocity resolution was 0.088 km s$^{-1}$ and the velocity coverage was 360 km s$^{-1}$ at 115 GHz. MOPS enabled simultaneous observations of \atom{C}{}{12}\atom{O}{}{}($J$=1--0), \atom{C}{}{13}\atom{O}{}{}($J$=1--0), and \atom{C}{}{}\atom{O}{}{18}($J$=1--0). The pointing accuracy was checked every 1 hour to keep within 7$''$ by observations of the 86 HGz SiO maser from AH Sco [R.A. (2000) = \timeform{17h11m17.16s}, Dec. (2000) = \timeform{-32D19'30.72''}]. The calibration from original output data onto $T^*_\mathrm{a}$ scale and the baseline fitting was done with the livedata task in AIPS++. The spectra were gridded to a 15$''$ spacing, and smoothed with a 36$''$ HPBW Gaussian function. We also smooth the channels in velocity to a 0.86 km s$^{-1}$ interval for \atom{C}{}{12}\atom{O}{}{}($J$=1--0) and \atom{C}{}{13}\atom{O}{}{}($J$=1--0) and a 2.0 km s$^{-1}$ for \atom{C}{}{}\atom{O}{}{18}($J$=1--0). The spectra were converted into a $T_{\mathrm{mb}}$ scale by dividing by "extended beam" efficiency 0.55 \citep{lad2005}. For comparison of \atom{C}{}{}\atom{O}{}{}($J$=1--0) with other excitation lines, the spectra were spatially smoothed with a 38$''$ Gaussian function beam, equivalent to the beam size of NANTEN2 at 460 GHz. We finally achieved the rms noise levels in the \atom{C}{}{12}\atom{O}{}{}($J$=1--0), \atom{C}{}{13}\atom{O}{}{}($J$=1--0), and \atom{C}{}{}\atom{O}{}{18}($J$=1--0) spectra of 0.13 K, 0.06 K, and 0.03 K, respectively (see also Table \ref{table:obsline}).

\subsection{ASTE observations}
 Observations of the \atom{C}{}{12}\atom{O}{}{}($J$=3--2) line were made using the ASTE (Atacama Submillimeter Telescope Experiment) 10m sub-mm telescope of NAOJ at Pampa La Bola at an altitude of 4800 m in Chile (Kohno 2005; Ezawa et al. 2004, 2008) for 7 days in August and September 2006. The HPBW was 22$''$ at 345 GHz. We used a position switching mode to cover the region shown in Figure \ref{fig:loop12.ii+sch}, in 1399 points, with a 40$''$ grid spacing. ASTE was equipped with 345 GHz SIS receiver SC 345 providing a typical system noise temperature of 190--300 K at 345 GHz for an elevation angle of 30--80 degrees in the double-side band (DSB). For checking the system stability and the absolute intensity calibration, M17SW [R.A. (1950) =\timeform{18h17m30.0s}, Dec. (1950) = \timeform{-16D13'6.0''}] was observed every 2--3 hours, the absolute temperature of which was assumed to be 69.6K \citep{wan1994}, and we adopt a beam efficiency of 0.6 at 345 GHz. The spectrometer comprises four digital back-end systems (autocorrelators) with 2048 channels. The total frequency bandwidth is 512 MHz, corresponding to a velocity coverage of 450 km s$^{-1}$ with a velocity resolution of 0.43 km s$^{-1}$ at 345GHz. We smoothed the data in velocity to a 0.86 km s$^{-1}$ resolution to improve the noise level and smoothed in space to a 38$''$ beam size for comparison with other excitation lines. The telescope pointing was measured to be accurate to within 2$''$ by radio observations of Jupiter and W Aql [R.A. (2000) = \timeform{19h15m23.21s}, Dec. (2000) = \timeform{-7D2'49.8''}]. Finally we got an RMS noise temperature level of $\sim$0.33 K in $T_{\mathrm{mb}}$.

\subsection{NANTEN2 observations}
 We used the NANTEN2 4m sub-mm telescope at 4800 m altitude at Pampa La Bola in Chile to observe the four regions in the foot points (Figure \ref{fig:lb.3-2.col}). These regions are chosen based on the results of \atom{C}{}{12}\atom{O}{}{}($J$=3--2) (see Section 3). Observations of the \atom{C}{}{12}\atom{O}{}{}($J$=4--3) transition at 460 GHz and the \atom{C}{}{12}\atom{O}{}{}($J$=7--6) transition at 810 GHz were made in June 2006 and December 2007. The HPBW in 460 GHz and 810 GHz were measured to be 38$''$ and 22$''$, respectively. The telescope was equipped with a dual-channel 460/810 GHz receiver. DSB receiver noise temperatures were $\sim$250 K at 460 GHz and $\sim$750 K at 810 GHz. The spectrometer was an acousto-optical spectrometer (AOS) with a bandwidth of 1 GHz of 2048 channels and the velocity resolution was 0.37 km s$^{-1}$ at 460 GHz and 0.21 km s$^{-1}$ at 806 GHz, respectively. We used the OTF mode with equatorial (J2000.0) coordinates and scanned in both RA and Dec directions for a 2$'$ $\times$ 2$'$ unit field. Main beam efficiencies at 460 GHz and 810 GHz were 0.5 and 0.45, respectively. The raw data were calibrated into $T^*_\mathrm{A}$ scale and then converted to $T_{\mathrm{mb}}$ scale by dividing by the main beam efficiency. The typical rms noise fluctuations were $\sim$0.45 K at 460 GHz and $\sim$0.8 K at 810 GHz at a velocity resolution of 0.86 km s$^{-1}$. For comparison of \atom{C}{}{12}\atom{O}{}{}($J$=7--6) with other lines, the spectra were smoothed to 38$''$.
 
\section{Results}
\subsection{Distribution of the molecular gas}
 Figure \ref{fig:lb.3-2.col} shows the integrated intensity distribution that includes the two foot points of loops 1 and 2 (Figure \ref{fig:lb.3-2.col}a) and part of the top of loop 1 (Figure \ref{fig:lb.3-2.col}b) in \atom{C}{}{12}\atom{O}{}{}($J$=3--2) at a 22$''\sim$0.9 pc resolution. The integration range in velocity, (with respect to the local standard of rest, LSR), is from $-180$ km s$^{-1}$ to $-40$ km s$^{-1}$. The foot points are generally elongated vertically to the plane and the most prominent feature ranges from $l=356.15^\circ$ to 356.25$^\circ$ and $b=0.8^\circ$ to 1.0$^\circ$. The rest of the distribution is extended with a few local maxima, e.g., toward ($l$, $b$)=(356.17$^\circ$, 0.75$^\circ$), (356.13$^\circ$, 0.78$^\circ$) and (356.25$^\circ$, 1.10$^\circ$). We have chosen four peaks in the foot points, A--D, and a peak in the loop top as listed in Table \ref{tab:obsotf} for a multi-J transition analysis in section 4. We note the emission is stronger in the eastern half of the distribution and that the eastern boundary shows a sharp intensity decrease perhaps barely resolved with the present resolution. Figures \ref{fig:lb.12.1-0.col}a and \ref{fig:lb.13.1-0.col}a show the distribution of \atom{C}{}{12}\atom{O}{}{}($J$=1--0) and \atom{C}{}{13}\atom{O}{}{}($J$=1--0) in the same velocity range at a 2 pc resolution with a spatial coverage limited to the bright part of \atom{C}{}{12}\atom{O}{}{}($J$=3--2) and these distributions are generally similar to the \atom{C}{}{12}\atom{O}{}{}($J$=3--2) distribution. Figures \ref{fig:lb.12.1-0.col}b and \ref{fig:lb.13.1-0.col}b show the different velocity range, $-70$ to $-10$ km s$^{-1}$, which exhibit a "U shape" (see section 3.2).
 
 \subsection{Velocity distribution}
  Figures \ref{fig:lvvb.3-2}--\ref{fig:lb.channel.12CO3-2.2} show the \atom{C}{}{12}\atom{O}{}{}($J$=3--2) velocity distribution of the foot points in two ways, i.e., position-velocity diagrams and velocity-channel distributions every 10 km s$^{-1}$, and Figures \ref{fig:lvvb.3-2.sch.eps} and \ref{fig:lb.channel.12CO3-2.2.sch.eps} show the labels and auxiliary lines superposed on the velocity distributions in order to show the components discussed below. 
 
  Figures \ref{fig:lvvb.3-2}a and \ref{fig:lvvb.3-2}b are latitude-velocity and longitude-velocity distributions of \atom{C}{}{12}\atom{O}{}{}($J$=3--2), respectively. The main part of the foot points, hereafter "main component", is distributed in a velocity range from $v$= $-150$ km s$^{-1}$ to $-40$ km s$^{-1}$ (Figure \ref{fig:lvvb.3-2.sch.eps}). The nature of the weaker emission for velocities greater than $-40$ km s$^{-1}$ was not yet discussed in F06, while the very narrow velocity feature at $\sim10$ km s$^{-1}$ is likely foreground outside the Galactic center. Examination of line intensity ratios like that of \atom{C}{}{12}\atom{O}{}{}($J$=3--2) to ($J$=1--0) indicates that most of the emission seen in Figure \ref{fig:lvvb.3-2} probably located in the Galactic center, as suggested by their high excitation (see section 4). We call the low velocity feature at $-30$ km s$^{-1}$ to 0 km s$^{-1}$ the "subcomponent" hereafter.

We note that the subcomponent is probably linked to the main component of the foot points, as suggested by the connecting broad emission at $b=0.8^\circ$ and by a few additional broad emission components at $b=0.88^\circ$, 0.94$^\circ$ and 1.0$^\circ$ (Figures \ref{fig:lvvb.3-2} and \ref{fig:lvvb.3-2.sch.eps}). The subcomponent shows a velocity gradient from $b=0.80^\circ$ to 0.90$^\circ$ in the opposite sense to that of the main component and forms a "U shape" in the $v$-$b$ diagram with the main component and the broad emission at $b=0.80^\circ$. Figure \ref{fig:lv.channel.+lb} shows the six latitude-velocity diagrams of the foot points every 0.03$^\circ$ in $l$. The main component and the subcomponent are clearly seen and appear linked at $b=1.0^\circ$ (a), $b=0.9^\circ$ (d,e) and $b=0.8^\circ$ (d,e,f). Below $b=0.8^\circ$ we also find another U shape in the $v$-$b$ diagram and name it U shape 2 which is distributed in $b=0.7^\circ$--0.8$^\circ$ and $v=-70$--$-10$ km s$^{-1}$ (Figures \ref{fig:lvvb.3-2} and \ref{fig:lvvb.3-2.sch.eps}). A protrusion toward the galactic plane is seen in $l=356.15^\circ$--356.20$^\circ$ and $b=0.7^\circ$--0.8$^\circ$ at $v=-70$--$-50$ km s$^{-1}$. In summary the distribution of the molecular gas consists of two U shapes and a protrusion as indicated in Figures \ref{fig:lb.channel.12CO3-2.2} and \ref{fig:lb.channel.12CO3-2.2.sch.eps}.
   
 Figure \ref{fig:lb.channel.12CO3-2.2} shows the velocity channel distributions of the main components in \atom{C}{}{12}\atom{O}{}{}($J$=3--2). The main component appears in every panel from $-120$ to $-40$ km s$^{-1}$ and subcomponent from $-30$ km s$^{-1}$ to 0 km s$^{-1}$. We recognize the intense part delineates a U shape in the bottom of the foot points as seen in panels from $-70$ km s$^{-1}$ to $-10$ km s$^{-1}$ (see also Figure \ref{fig:lb.channel.12CO3-2.2.sch.eps}). This U shape is also seen in $^{12}$CO($J$=1--0) (Figures \ref{fig:lb.12.1-0.col}b and \ref{fig:lb.13.1-0.col}b around $b$=0.8$^\circ$). 
 
 Peaks A--D are seen over a broad velocity range. Peak A is seen in two panels in Figure \ref{fig:lb.channel.12CO3-2.2} from $-110$ to $-90$ km s$^{-1}$, peak B in four panels from $-80$ to $-40$ km s$^{-1}$, peak C in four panels from $-100$ to $-60$ km s$^{-1}$, peak D in four panels from $-80$ to $-40$ km s$^{-1}$. The most intense region in the foot point is peak C, which is connected to the bottom of the U shape, and peak D corresponds to the protrusion and shows a sharp inverse-triangle like shape. Peak D also seems to be connected to the component at $l=$356.06$^\circ$ and $b=$0.76$^\circ$ at $v=-10$ km s$^{-1}$. These configure the another U shape. 
Peak B corresponds to one of the broad features shown in Figures \ref{fig:lvvb.3-2} and \ref{fig:lvvb.3-2.sch.eps}. Other two broad features can be seen at $l=356.25^\circ$ and $b=0.90^\circ$ to 1.10$^\circ$ at velocity from $-60$ to $-20$ km s$^{-1}$ and at $l=356.08^\circ$ and $b=0.76^\circ$ to 0.80$^\circ$ (Figure \ref{fig:lb.channel.12CO3-2.2.sch.eps}). These two features were not observed except in $^{12}$CO($J$=3--2) emission.
 A general trend in Figure \ref{fig:lb.channel.12CO3-2.2} is that the peaks A, C and D show sharp intensity decreases toward the east and south with weaker "tails" toward the north. Figure \ref{fig:A3_3.peakV.gt75.small2.13.col} shows the line profile of peak C where the \atom{C}{}{12}\atom{O}{}{} peak velocity changes by $\sim10$ km s$^{-1}$ every $\sim$20 pc with a decrease of the peak intensity by a factor of 4 toward the plane. Such a trend is not clearly seen for peak B, embedded in spatially extended emission in the panels from $-70$ to $-60$ km s$^{-1}$. 
   
\subsection{Spectra presentation and line intensity ratios}
  We show the line spectra obtained at the four peaks A--D as well as at the peak position at the top of the loop in Figure \ref{fig:specall.a-d.new}. All the spectra are smoothed to a 38$''$ Gaussian beam for comparison with the \atom{C}{}{12}\atom{O}{}{}($J$=4--3) profiles and are smoothed to 0.86 km s$^{-1}$ resolution. Only the \atom{C}{}{}\atom{O}{}{18}($J$=1--0) spectra were smoothed to 2.0 km s$^{-1}$ resolution to obtain better signal-noise ratios. The \atom{C}{}{12}\atom{O}{}{}($J$=1--0, 3--2, 4--3) and \atom{C}{}{13}\atom{O}{}{} ($J$=1--0) transitions are clearly detected while the \atom{C}{}{12}\atom{O}{}{} ($J$=7--6) transition is not detected, with an upper limit of 0.76 K. The line widths of all the spectra are broad with velocity extents of more than 20--40 km s$^{-1}$.
    
 We investigate the intensity ratios, $R_{j/i}$, in order to reveal the excitation condition of the gas. Here $R_{j/i}$ stands for an intensity ratio of emissions from lien $i$ to line $j$. $^{12}$CO($J$=1--0), $^{12}$CO($J$=3--2), $^{12}$CO($J$=4--3), $^{12}$CO($J$=7--6), $^{13}$CO($J$=1--0) and C$^{18}$O($J$=1--0) for $i$ and $j$ are represented as 1--0, 3--2, 4--3, 7--6, 13 and 18, respectively.
 We show three histograms of line intensity ratios in the four velocity ranges of $-140$--$-40$ km s$^{-1}$ (white), $-40$--$-20$ km s$^{-1}$ (red), $-20$--0 km s$^{-1}$ (orange) and 0--20 km s$^{-1}$ (green) and the combinations of these velocity ranges (black) in Figure \ref{fig:histo.ratio.3-2.1-0.13.2}. Figure \ref{fig:histo.ratio.3-2.1-0.13.2}a shows $R_{3-2/1-0}$; this indicates that the loop emission is characterized by high ratios peaked at 0.6 with a range from 0.3 to 0.95 at a 90 \% peak level, somewhat smaller than $R_{3-2/1-0}$ of 0.9 in CMZ estimated by \citet{oka2007}. The local emission has a narrow linewidths mostly in a range from 0 to 20 km s$^{-1}$ is peaked at 0.2 with a 90 \% range of 0.1 to 0.6. The other two show peaks near 0.6 while the $-20$--0 km s$^{-1}$ component shows another peak at 0.3 similar to the local clouds. The other two histograms, $R_{1-0/13}$, (Figure \ref{fig:histo.ratio.3-2.1-0.13.2}b) and $R_{3-2/13}$ (Figure \ref{fig:histo.ratio.3-2.1-0.13.2}c), also indicate that the loop component and the components in a range from $-40$ to 0 km s$^{-1}$ show ratios distinct from the local component from 0 to 20 km s$^{-1}$. Figure \ref{fig:histo.tau} shows a histogram of the \atom{C}{}{12}\atom{O}{}{} optical depth estimated by taking ratio $R_{1-0/13}$ with assumption of the abundance ratio [\atom{C}{}{12}]/[\atom{C}{}{13}]. 
\citet{riq2010} estimated [$^{12}$C]/[$^{13}$C] in the loops as $\sim$50--70. This is a typical one in the inner Galaxy and is higher than 24 estimated in the CMZ (i.e. \cite{lan1990}). Here we shall assume the abundance ratio [\atom{C}{}{12}]/[\atom{C}{}{13}] $\sim$ 53 estimated by \citet{wil1994}. Figure \ref{fig:histo.tau} indicates that the gas in the loops tends to show smaller optical depths peaked at around 3--4, while the local component of $-20$ km s$^{-1}$ is peaked at around 6--10. This estimate is similar to the optical depth derived in the CMZ by \citet{oka1998}.
We summarize that the loop component is best characterized by high excitation conditions as indicated by the high ratio of the \atom{C}{}{12}\atom{O}{}{}($J$=3--2) to \atom{C}{}{12}\atom{O}{}{}($J$=1--0) line intensities and the smaller \atom{C}{}{12}\atom{O}{}{} optical depth. 
 
 Figure \ref{fig:ratio.channel.lb.3-2.1-0.2} show the distributions of the ratio $R_{3-2/1-0}$ in velocity channel distributions. Generally speaking, the ratio is correlated with the \atom{C}{}{12}\atom{O}{}{}($J$=3--2) intensity and the ratio becomes higher than 0.7 when the \atom{C}{}{12}\atom{O}{}{}($J$=3--2) intensity is higher than $\sim$44 K km s$^{-1}$. The most notable enhanced ratio, corresponding to peak B (broad emission), is found in Figure \ref{fig:ratio.channel.lb.3-2.1-0.2} at $(l, b)\sim(356.18^\circ, 0.92^\circ)$ for velocity range of $-80$--$-10$ km s$^{-1}$, where the highest ratio is around 2.0. The secondary enhancement is seen at $(l, b)\sim(356.18^\circ, 0.7^\circ$--$0.8^\circ)$ and velocity range of $-80$--$-60$ km s$^{-1}$, showing a ratio around 1.0 and corresponding to peak D.  

 Figure \ref{fig:ratio.channel.vb.3-2.1-0.2} shows $v$-$b$ diagrams of the $R_{3-2/1-0}$ and is generally consistent with Figure \ref{fig:ratio.channel.lb.3-2.1-0.2}. The most enhanced component around peak B shown in Figure \ref{fig:ratio.channel.lb.3-2.1-0.2} is found toward the broad feature at $b$ $\sim$ 0.9$^\circ$--0.95$^\circ$ for longitude range of 356.17$^\circ$--356.20$^\circ$. In addition, a high ratio feature with a very narrow velocity width of 1--2 km s$^{-1}$ is seen at $v\sim-80$ km s$^{-1}$ and $l=356.2^\circ$ with a weak velocity gradient from $b=0.85^\circ$ to 1.05$^\circ$ and another similar feature is seen at $v\sim-30$ km s$^{-1}$ and $l=356.15^\circ$ from $b=0.80^\circ$ to 0.95$^\circ$. These narrow features are smeared out in Figure \ref{fig:ratio.channel.lb.3-2.1-0.2}.
 
\section{Data analysis}
\subsection{Physical properties; Clumps}
We identified clumps in the following way in the \atom{C}{}{12}\atom{O}{}{}($J$=3--2) integrated intensity distributions shown in Figure \ref{fig:lb.3-2.col}; 1) Find local peaks that have the integrated intensity stronger than half of that in peak C, because peak C shows the maximum integrated intensity level in the foot point. 2) Draw a contour at two thirds of the peak integrated intensity level and identify it as a clump unless it contains other local peaks, 3) If there are other peaks inside the contour, draw another contour with interval of 10 $\sigma$ and find a contour which has no other peaks inside, and 4) If we find multiple peaks with an isolated contour when we perform the operation (3), we should identify these all peaks as local peaks and repeat operations (2) and (3).
 
In this way, we identified six clumps in the foot point. Then, we derived physical properties of the clumps. The results are summarized in Table \ref{table:clumpphys}. The radii of the clumps, $r$, are $\sim$3--6 pc, and the intensity weighted velocity dispersions (FWHM), $\Delta V$, at the peak positions are $\sim$20--36 km s$^{-1}$. We estimate the clump masses in two ways; the mass estimated from $r$ and $\Delta V$ by assuming virial equilibrium, $M_\mathrm{vir}$, and the mass estimated from \atom{C}{}{13}\atom{O}{}{} by assuming the local thermodynamic equilibrium (LTE). $M_\mathrm{vir}$ is calculated as follows by assuming uniform density distribution;
\begin{eqnarray}
       M_\mathrm{vir} = 209 \left( \frac{r}{\mathrm{pc}} \right) \left( \frac{\Delta V}{\mathrm{km \ s^{-1}}} \right)^2 \ \MO
 \end{eqnarray}
\atom{C}{}{13}\atom{O}{}{} column densities are calculated by assuming the LTE condition to estimate the molecular mass. The optical depth of \atom{C}{}{13}\atom{O}{}{}, $\tau(\mathrm{\atom{C}{}{13}\atom{O}{}{}})$, was calculated using following equation;
\begin{eqnarray}
       \tau(\mathrm{\atom{C}{}{13}\atom{O}{}{}}) = - \ln \left[ 1 - \frac{T^{*}_{\mathrm{R}}(\mathrm{\atom{C}{}{13}\atom{O}{}{}})}{5.29 \times (J(T_{\mathrm{ex}})-0.164)} \right]
 \end{eqnarray}
where, $T^{*}_{\mathrm{R}}(\mathrm{\atom{C}{}{13}\atom{O}{}{}})$ and $T_{\mathrm{ex}}$ are the radiation temperature and the excitation temperature of \atom{C}{}{13}\atom{O}{}{}, respectively. $J(T)$ is defined as $J(T) = 1 / \left[ \exp(5.29/T)-1 \right]$. $N(\mathrm{\atom{C}{}{13}\atom{O}{}{}})$ was estimated from:
\begin{eqnarray}
       N(\mathrm{\atom{C}{}{13}\atom{O}{}{}}) = 2.42 \times 10^{14} \frac{\tau(\mathrm{\atom{C}{}{13}\atom{O}{}{}}) \Delta V T_{\mathrm{ex}}}{1-\exp(-5.29/T_{\mathrm{ex}})} \ (\mathrm{cm}^{-2}).
 \end{eqnarray}
In this study, we assume uniform $T_{\mathrm{ex}}$ of 40 K as discussed in the next subsection. We assume two abundance ratios [H$_2$]/[\atom{C}{}{13}\atom{O}{}{}] = 10$^6$, the value of Sgr B2 \citep{lis1989}, and $5\times10^5$, the average value of local clouds \citep{dic1978}, to convert $N$(\atom{C}{}{13}\atom{O}{}{}) into $N$(H$_2$). $M_\mathrm{CO}$ of each clump is estimated by using 
\begin{eqnarray}
      M_{\atom{C}{}{}\atom{O}{}{}} = 2.8m_{\atom{H}{}{}} \sum \left[ D^2\Omega N(\atom{H}{}{}_2) \right] \ \MO.
 \end{eqnarray}
while $M_\mathrm{vir}$ is in the order of $\sim$10$^5$--10$^6$ \MO, $M_{\atom{C}{}{}\atom{O}{}{}}$ is only in the order of $10^4$ \MO. It means that the dynamical state of the molecular gas is considerably different from the virial equilibrium. 
If we assume the uniform velocity dispersion of 20 km s$^{-1}$ as a lower limit for the turbulent velocity, the kinetic energy of each clumps is estimated to $\sim10^{50}$ erg.

\subsection{Physical properties; Temperature and Density}
 We applied the large velocity gradient (LVG) analysis (\cite{gol1974}; \cite{sco1974}) to estimate the physical parameters of the molecular gas toward the loop foot points and the loop top by adopting a spherically symmetric uniform model having a radial velocity gradient $dv/dr$. We calculate level populations of \atom{C}{}{12}\atom{O}{}{}, \atom{C}{}{13}\atom{O}{}{} and \atom{C}{}{}\atom{O}{}{18} molecular rotational states and line intensities under these assumptions. The LVG model requires three independent parameters to calculate emission line intensities; kinetic temperature, $T_\mathrm{k}$, density of molecular hydrogen, $n(\atom{H}{}{}_2)$, and $X(\atom{C}{}{}\atom{O}{}{})/(dv/dr)$. $X(\atom{C}{}{}\atom{O}{}{})/(dv/dr)$ is the abundance ratio of \atom{C}{}{}\atom{O}{}{} to \atom{H}{}{}$_2$ divided by the velocity gradient in the cloud. 
We use the abundance ratios [\atom{C}{}{12}]/[\atom{C}{}{13}] $\sim$ 53 and [\atom{O}{}{16}]/[\atom{O}{}{18}] $\sim$ 327 \citep{wil1994} and the molecular abundance [\atom{C}{}{12}\atom{O}{}{}]/[\atom{H}{}{}$_2$] $\sim$ $10^{-4}$ as a typical value for the inner Galaxy (i.e. \cite{fre1982,leu1984,bla1987}).
We estimate the mean velocity gradient within the foot point as $\sim$9.0 km s$^{-1}$ pc$^{-1}$ (Table \ref{table:clumpphys}) and $X(\atom{C}{}{}\atom{O}{}{})/(dv/dr)$ to be $1.1 \times 10^{-5}$ for \atom{C}{}{12}\atom{O}{}{}. 

In order to solve for the temperatures and densities which reproduce the observed line intensity ratio, we calculate chi-square defined as below;
\begin{eqnarray}
\chi^2 = \sum^{N-1}_{i=1}\sum^N_{j=i+1}\left[\frac{\{ R_\mathrm{obs}(i, j) - R_\mathrm{LVG}(i, j) \}^2}{\sigma(i, j)^2}\right]
\end{eqnarray}
where $N$ is the number of transitions of the observed molecule. $i$ and $j$ refer to different molecular transitions, $R_\mathrm{obs}(i, j)$ is the observed line intensity ratio from transition $i$ to transition $j$ and $R_\mathrm{LVG}(i, j)$ is the ratio between transitions $i$ and $j$ estimated from the LVG calculations. The standard deviation $\sigma(i, j)$ for $R_\mathrm{obs}(i, j)$ is estimated by considering the noise level of the observations and the calibration error.
We assume that the error of calibration from $T^*_\mathrm{a}$ to $T_{\mathrm{mb}}$ is uniformly 10 \% for all observations. Because we use the same calibration factor for the spectra obtained with Mopra, the relative calibration error has no effect when we take a ratio between intensities obtained with Mopra; i.e., \atom{C}{}{12}\atom{O}{}{}($J$=1--0),  \atom{C}{}{13}\atom{O}{}{}($J$=1--0) and \atom{C}{}{}\atom{O}{}{18}($J$=1--0). In order to reduce the error, we use the average line intensities over 10 km s$^{-1}$ bins except for \atom{C}{}{}\atom{O}{}{18}, because \atom{C}{}{}\atom{O}{}{18} has already been smoothed to be 2.0 km s$^{-1}$ velocity resolution as discussed in section 3.3 and Figure \ref{fig:specall.a-d.new}, further smoothing did not give a better noise level.
The degree of freedom, $\nu$, is defined as follows; 
\begin{eqnarray}
\nu = \left( \begin{array}{c} N\\ 2\\ \end{array} \right)
\end{eqnarray}
The data used here are derived from the line profiles in Figure \ref{fig:specall.a-d.new} and the ratios are estimated for peaks A--D and the loop top with all transitions. 
Therefore, the degree of freedom is given as 15 for N$=6$. The intensities of all transitions used here are listed in Table \ref{tab:ratio}. Because we could not detect any emissions of $^{12}$CO($J$=7--6), we use the 1$\sigma$ noise levels for calculations as upper limit. Since $\chi^2$ is weighted by $\sigma^2$, this assumption does not affect the results.

Figure \ref{fig:lvg.ratio_plot} shows the loci of constant $R_{j/i}$ for the five $R_{j/i}$'s as a function of density and temperature.
We find $R_{3-2/1-0}$ is sensitive to a large density range from 10$^2$ cm$^{-3}$ to 10$^5$ cm$^{-3}$ while $R_{4-3/1-0}$ is sensitive to higher density above $10^3$ cm$^{-3}$. We also note that  $R_{1-0/13}$ is nearly orthogonal to $R_{3-2/1-0}$ for density lower than 10$^4$ cm$^{-3}$, making the combination useful in obtaining solutions. 

Figures \ref{fig:LVGall.peakA}--\ref{fig:LVGall.peakTOP} and Table \ref{tab:lvg} show the results of fitting the data obtained with a $\chi^2$ minimization approach to find the best solution of temperature and density. Each thick locus surrounding the cross indicates the $\chi^2$ of 25.0, which corresponds to the 5\% confidence level of $\chi^2$ distribution with 15 degree of freedom. The crosses denote the lowest point of $\chi^2$. Three intensity ratios, $R_{3-2/1-0}$, $R_{4-3/1-0}$ and $R_{1-0/13}$, are also shown by thin lines. The $-75$ km s$^{-1}$ and $-15$ km s$^{-1}$ components at peak B do not have any solutions with 5\% confidence level. Another presentation of the results is found in Figure \ref{fig:specall.a-d.+LVG} for the five peaks along with the \atom{C}{}{12}\atom{O}{}{}($J$=3--2) line profiles. An error bar in the figure is defined as 5\% confidence level of $\chi^2$ distribution. The density is in the range from 10$^3$ to 10$^4$ cm$^{-3}$ and the temperature is from $\sim30$--40 K to 100 K or higher. As expected from the $R_{3-2/1-0}$ values in Figures \ref{fig:ratio.channel.lb.3-2.1-0.2} and \ref{fig:ratio.channel.vb.3-2.1-0.2}, the broad emission connecting different sides of the U shape (peak B) shows the highest temperature, 100 K or higher, amongst the five peaks.

\section{Discussion}
There are no candidates for protostars identified by IRAS in the observed regions. The 10 GHz radio continuum emission observed by \citet{han1987} gives an upper limit of 0.1 K. This 10 GHz flux density is equivalent to 10$^{43.5}$ FUV photons s$^{-1}$, assuming 220 square arc minutes as the area of the foot point and 8000 K electron temperature. That corresponds to a single B3 star or a star of later spectral type (\cite{kur1994}; \cite{pan1973}; \cite{mez1967}). Only one ultra compact \atom{H}{}{}\emissiontype{II} region (UC\atom{H}{}{}\emissiontype{II}) is identified at ($l$, $b$) $\sim$ (356.25$^\circ$, 0.7$^\circ$) at a velocity of $+120$ km s$^{-1}$ from \atom{H}{}{}$\alpha$ recombination line observations (\cite{cas1987}; \cite{loc1989}). So, the UC\atom{H}{}{}\emissiontype{II} is not associated with the foot point. Here, there is no indication of significant formation of massive stars. Some other mechanism for heating the molecular gas is required, in particular, towards peaks B and C. 

 The magnetic flotation model presented by F06 is able to offer another explanation to heat the gas by shocks. F06 suggested that loops 1 and 2 are created by magnetic buoyancy driven by the Parker instability. The floated gas falls down to the Galactic plane along the magnetic loop, and the accumulated gas forms a massive cloud in the foot point of the loop. In this case, if the fall down velocity of the gas exceeds the speed of sound, shock fronts are formed in the foot point \citep{mat1988}. The magnetic field is estimated to be 150 $\mu$G in the loops under the assumption of energy equipartition, and the velocity of the gas is estimated to be similar to the Alfv\'en speed, $\sim24$ km s$^{-1}$ (F06). 
 
 The structure of shock waves in molecular clouds is an important subject. \citet{dra1983} made such calculations including the effects of ion-neutral streaming driven by the magnetic field. They found that shock waves in molecular clouds are usually C-type shock waves, mediated by the dissipation accompanying ion-neutral streaming, and in which all of the hydrodynamic variables are continuous. In the foot point of the loops, C-type shock seems to be a viable model because they occur with a strong magnetic field and at a moderate shock speed. The limiting shock speed for C-type shocks not dissociating the \atom{H}{}{}$_2$ molecules is estimated to be 45 km s$^{-1}$ \citep{dra1983}. The Alfv\'en speed 24 km s$^{-1}$ estimated by F06 and the velocity dispersion of the clumps shown in Table \ref{table:clumpphys} are both within this range and hence we adopt C-type shock as the heating mechanism for the loops.

\citet{dra1983} show that the neutral gas temperature is in a range 100--1000 K for a density of 10$^4$ cm$^{-3}$, $B$ of 100 $\mu$G and shock speed of 20--40 km s$^{-1}$. These parameters are consistent with density of 10$^3$--10$^4$ cm$^{-3}$, magnetic field of 150 $\mu$G and a falling speed of 24 km s$^{-1}$ (F06). Such high temperature is consistent with the present estimate of temperature higher than 30--40 K, since the high temperature layer should be thin like $10^{16}$--10$^{17}$ cm, $\sim$10--100 times smaller than the current resolution of a few pc. \citet{chi1998} also presented results on shock parameters consistent with the above.

We have compared the present results with local two-dimensional MHD numerical simulations (\cite{mat1988}; \cite{tak2009}). The distribution of the foot point is basically characterized by a U shape both in space (section 3.2 and Figures \ref{fig:lb.12.1-0.col}b, \ref{fig:lb.13.1-0.col}b and \ref{fig:lb.channel.12CO3-2.2}) and in velocity (section 3.2 and Figures \ref{fig:lvvb.3-2} and \ref{fig:lv.channel.+lb}). MHD simulations indeed indicate that the foot point between two loops becomes U shaped, consisting of the two down flows from the two loops that merges at the bottom (see e.g., Figures 4 in \cite{tak2009}). In velocity space, the foot point foreground and background components have a large velocity splitting of order the Alfv\'en speed. It is preferable to compare these observational properties with the MHD calculations to confirm the scenario. There is, however, no previous study which focuses on the details in the foot point of the galactic loops. In the actual formation of loops, we need to take into account a time lag in falling motion between the two sides of a foot point as well as the other physical differences between them. In the MHD calculations, such conditions are simply assumed to be uniform. It is neither shown yet by MHD calculations how the compact broad features connecting the U shape are formed (Figures 8 and 9). In order to better understand the whole process, we must await detailed numerical simulations of the foot point formation with take into account the observed parameters and also time evolution.
 
 The present analysis indicates that the highest ratio in the $R_{3-2/1-0}$ is formed not toward the bottom of the foot points but well above the foot points apparently separated from the strongly shocked layer in the bottom of the U shape. It is interesting to compare these results with the solar foot points for which a number of observations have been accumulated on magnetic activity. 
 
 \citet{iso2007} argued that anti-parallel field lines are produced in the foot point when multiple loops rose. This causes magnetic reconnection that can lead to a jet-like structure rising upward. Or, alternatively, in the U shaped region the field lines become compressed to reduce the falling gas flux, leading to gas bouncing at the narrow neck \citep{kud1999}. We present a schematic figure of these scenarios in Figure \ref{fig:reconnection.sch}. Numerical simulations in the solar case suggest that the initial condition may determine which process becomes dominant, either reconnection or bouncing. The basic physics is common to the spurs in Galactic loops as already discussed by \citet{mat1988} and \citet{tak2009}. 

We shall make a crude estimate of the energies concerned, both the magnetic and gravitational energies released. Here we assume the concerned volume of the U shape as a cylinder with a radius of 10 pc and a height of 30 pc. Gravitational energy is released by the gas falling down along the loop that can lead to shock heating of the U shape. The heights of the tops of loops 1 and 2 and the U shape from the galactic plane are estimated as $\sim$200 pc, $\sim$300 pc and $\sim$130 pc, respectively, at a distance to the Galactic center of 8.5 kpc, and the total molecular mass in the U shape is estimated as $\sim5\times10^4$ \MO. Potential energies from the U shape to the tops of loops 1 and 2 divided by the time scale of 10$^7$ years (Machida et al. 2009; Takahashi et al. 2009) at the galacto-centric radius of 670 pc are derived as $0.5\times10^{37}$ erg s$^{-1}$ and $1.3\times10^{37}$ erg s$^{-1}$, respectively, by using the modified Miyamoto-Nagai potential \citep{miy1975,sof1996}. If the reconnection takes place, we need to take into account the additional magnetic energy release. We use the magnetic field strength and Alfv\'en speed of 150 $\mu$G and 24 km s$^{-1}$, respectively, estimated by F06. According to the scenario given above, the field lines move parallel to the galactic plane and inflow into the U shape. If we roughly assume the inflow speed to be 10\% of the Alfv\'en speed, the total magnetic energy accumulated into the cylinder is estimated to $\sim4\times10^{50}$ erg in $\sim$10$^7$ years. This energy is comparable to the magnetic energy of $\sim10^{51}$ ergs for a typical single loop given by \citet{mac2009}. In $\sim10^6$ years after the magnetic reconnection occurs, the maximum available power of the reconnection is derived as $\sim1.3\times10^{37}$ erg s$^{-1}$. Thus, the total available energy injected to the U shape is $1.8\times10^{37}$--$2.6\times10^{37}$ erg s$^{-1}$ by summing up both the magnetic and gravitational energies. The cooling power in peak B is estimated to be $\sim1.7\times10^{36}$ erg s$^{-1}$ for kinetic temperature of 100 K and density of $4\times10^3$ cm$^{-3}$ for a uniform sphere with $\sim$3 pc radius and that over the entire U shape is estimated to be  $\sim4.3\times10^{36}$ erg s$^{-1}$ for 40 K and $4\times10^3$ cm$^{-3}$ \citep{gol1978}. These values correspond to nearly comparable to, or $\sim$20\% of, the maximum available power. We therefore infer that the reconnection is able to explain the heating of the warm gas in the U shape, while obviously we require more detailed model simulations and direct measurements of the magnetic field to reach a firm conclusion on the physical processes related to the heating mechanism.
 
\section{Conclusions}
We have made detailed high-resolution observations of the foot points of the molecular loops 1 and 2 \citep{fuk2006} in six rotational transitions of the interstellar CO molecule. The main conclusions are summarized below;

\begin{figure}
  \begin{center}
    \FigureFile(127.95mm,112.83mm){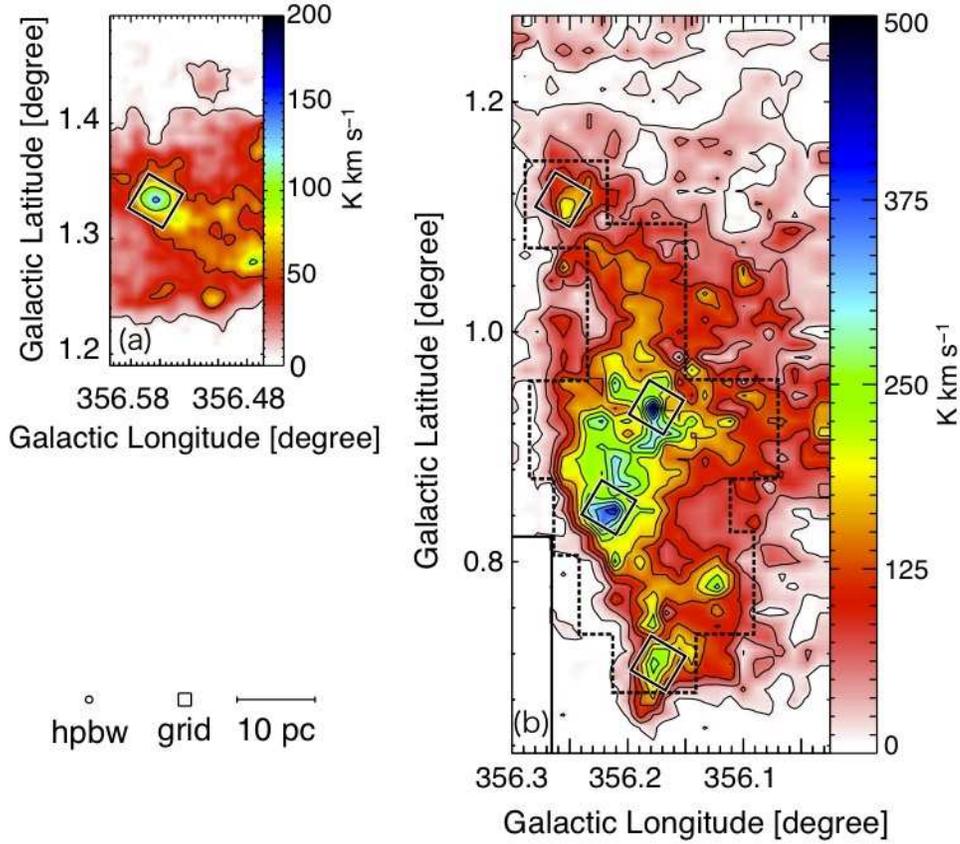}
  \end{center}
  \caption{Integrated intensity distributions of $^{12}$CO($J$=3--2) emissions at the top and foot point of the loops. (a) Top of the loop1. Integration range is from $-180$ to $-40$ km s$^{-1}$. Contours are plotted every 40 K km s$^{-1}$ from 15 K km s$^{-1}$. (b) Foot point of the loops. Integration range and contour levels are the same as that in (a). The dotted box shows the observing region of $^{12}$CO($J$=1--0) emissions and $^{13}$CO($J$=1--0) emissions, that are shown in Figures \ref{fig:lb.12.1-0.col} and \ref{fig:lb.13.1-0.col}. The small boxes show the observed reigons of $^{12}$CO($J$=4--3, 7--6) emissions. }\label{fig:lb.3-2.col}
\end{figure}

\begin{figure}
  \begin{center}
    \FigureFile(135.93mm,91.18mm){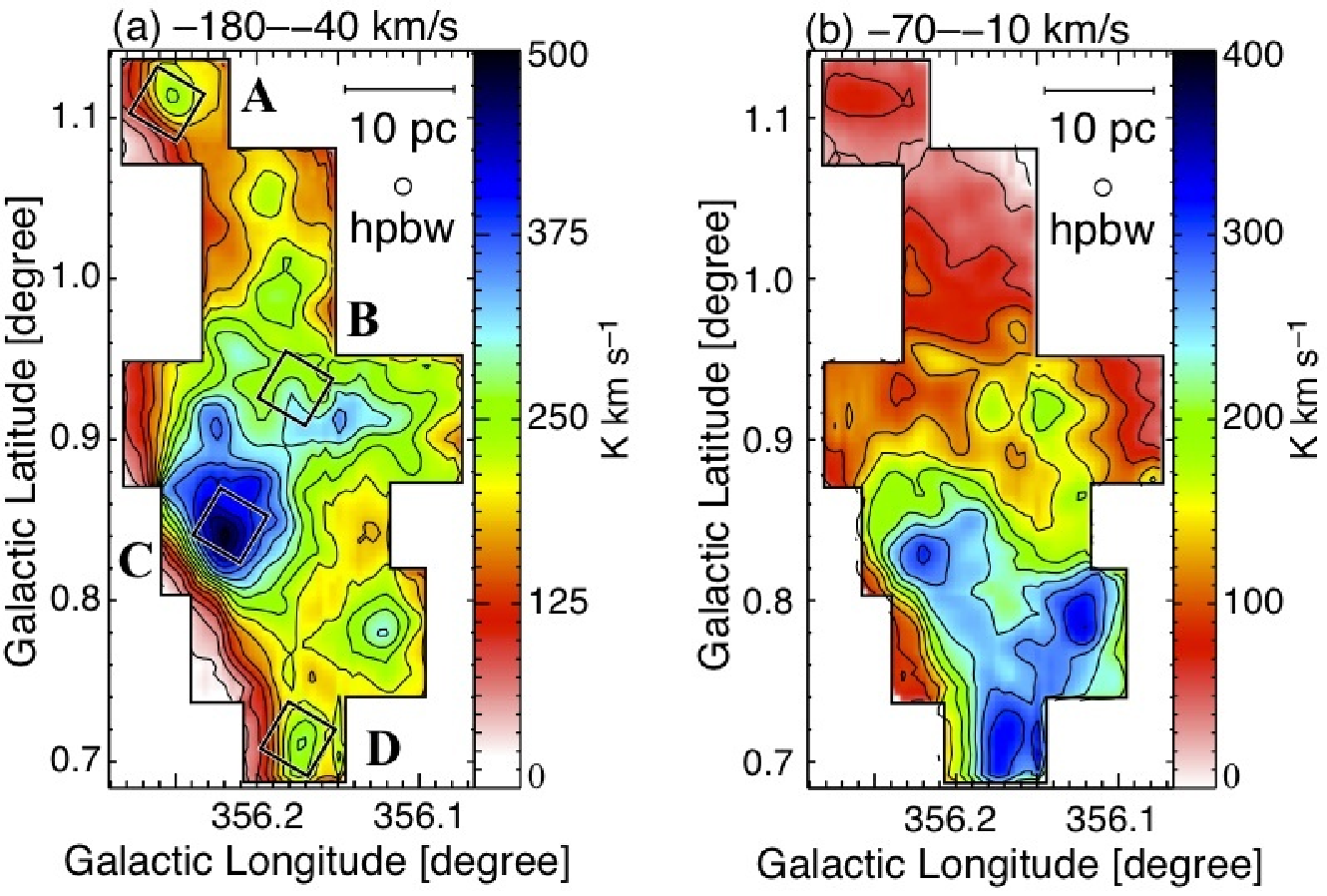}
  \end{center}
  \caption{Integrated intensity distribution of $^{12}$CO($J$=1--0) emissions at the foot point of the loops. Integration ranges are from $-180$ to $-40$ km s$^{-1}$(a) and from $-70$ to $-10$ km s$^{-1}$(b). Contours are plotted every 30 K km s$^{-1}$ from 2 K km s$^{-1}$. The labeled clumps are discussed in the text.}\label{fig:lb.12.1-0.col}
\end{figure}

\begin{figure}
  \begin{center}
    \FigureFile(133.81mm,91.18mm){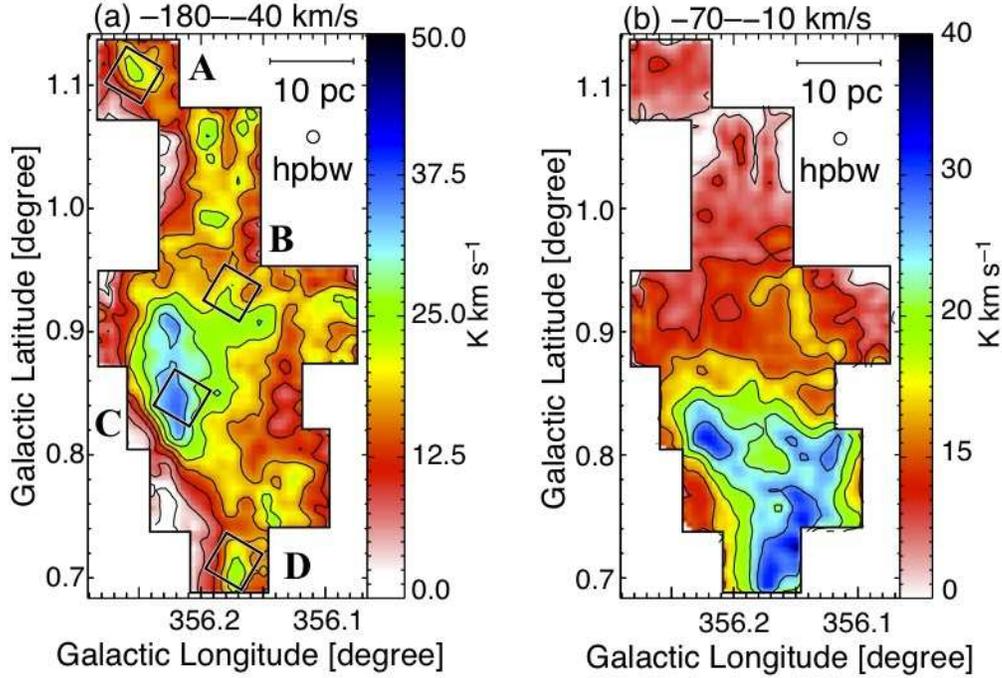}
  \end{center}
  \caption{Integrated intensity distribution of \atom{C}{}{13}\atom{O}{}{}($J$=1--0) emissions at the foot point of the loops. Integration ranges are from $-180$ to $-40$ km s$^{-1}$(a) and from $-70$ to $-10$ km s$^{-1}$(b). Contours are plotted every 5 K km s$^{-1}$ from 2 K km s$^{-1}$. }\label{fig:lb.13.1-0.col}
\end{figure}

\begin{figure}
  \begin{center}
    \FigureFile(139.11mm,71.35mm){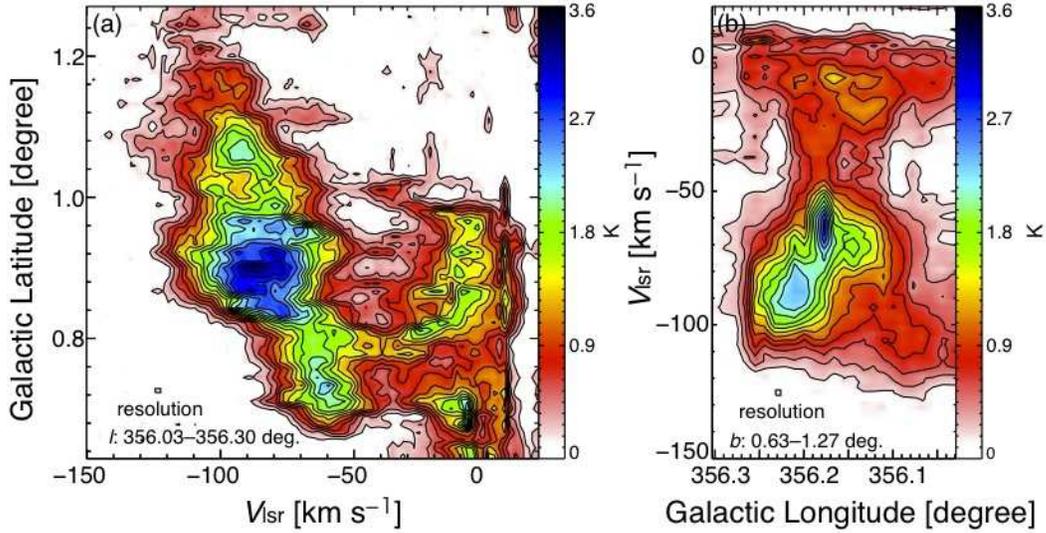}
  \end{center}
  \caption{(a) Velocity - galactic latitude diagram of the footpoint of the loops in \atom{C}{}{12}\atom{O}{}{}($J$=3--2) averaged from 356.03$^\circ$ to 356.27$^\circ$ in galactic longitude. Contours are plotted every 0.2 K. (b) Galactic longitude-velocity diagram of the footpoint of the loops in \atom{C}{}{12}\atom{O}{}{}($J$=3--2) averaged from 0.63$^\circ$ to 1.27$^\circ$  in galactic latitude. Contours are plotted every 0.2 K.}\label{fig:lvvb.3-2}
\end{figure}

\begin{figure}
  \begin{center}
    \FigureFile(133.25mm, 96.65mm){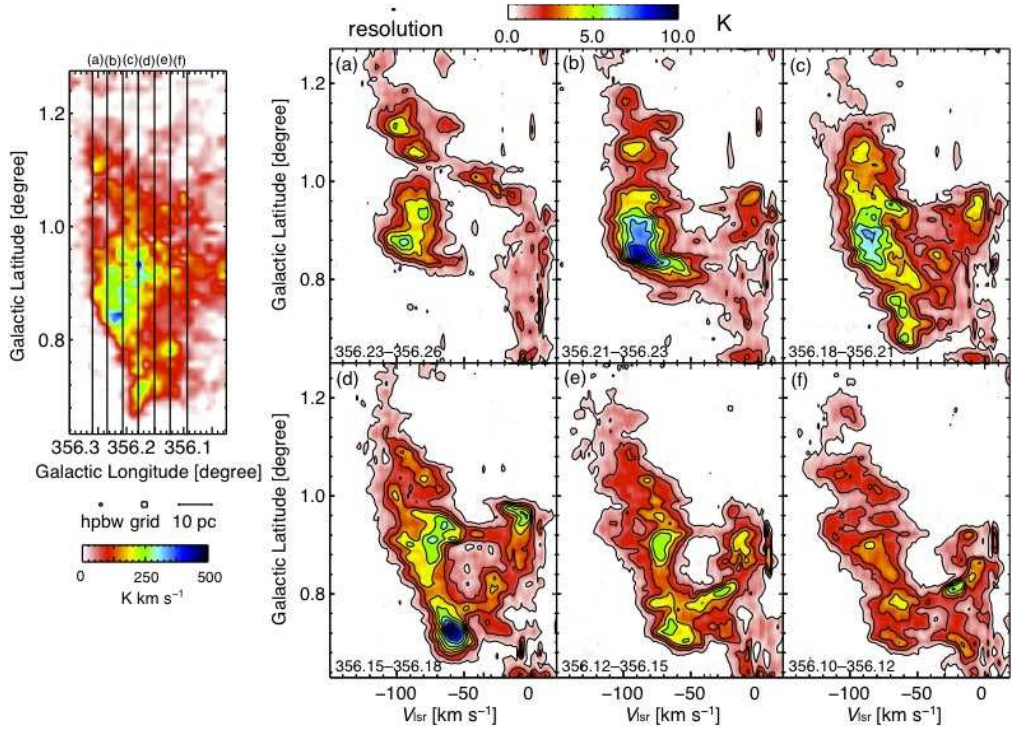}
  \end{center}
  \caption{(a--f) Longitude channel maps of $^{12}$CO($J$=3--2) averaged over successive 100$''$ intervals. Contours are plotted every 1 K from 0.5 K. The figure in the left side is the integrated intensity distributions which is the same as the image shown in figure \ref{fig:lb.3-2.col}. Solid lines in the figure show the integration ranges of figures (a)--(f).}\label{fig:lv.channel.+lb}
\end{figure}

\begin{figure}
  \begin{center}
    \FigureFile(129.85mm, 184.99mm){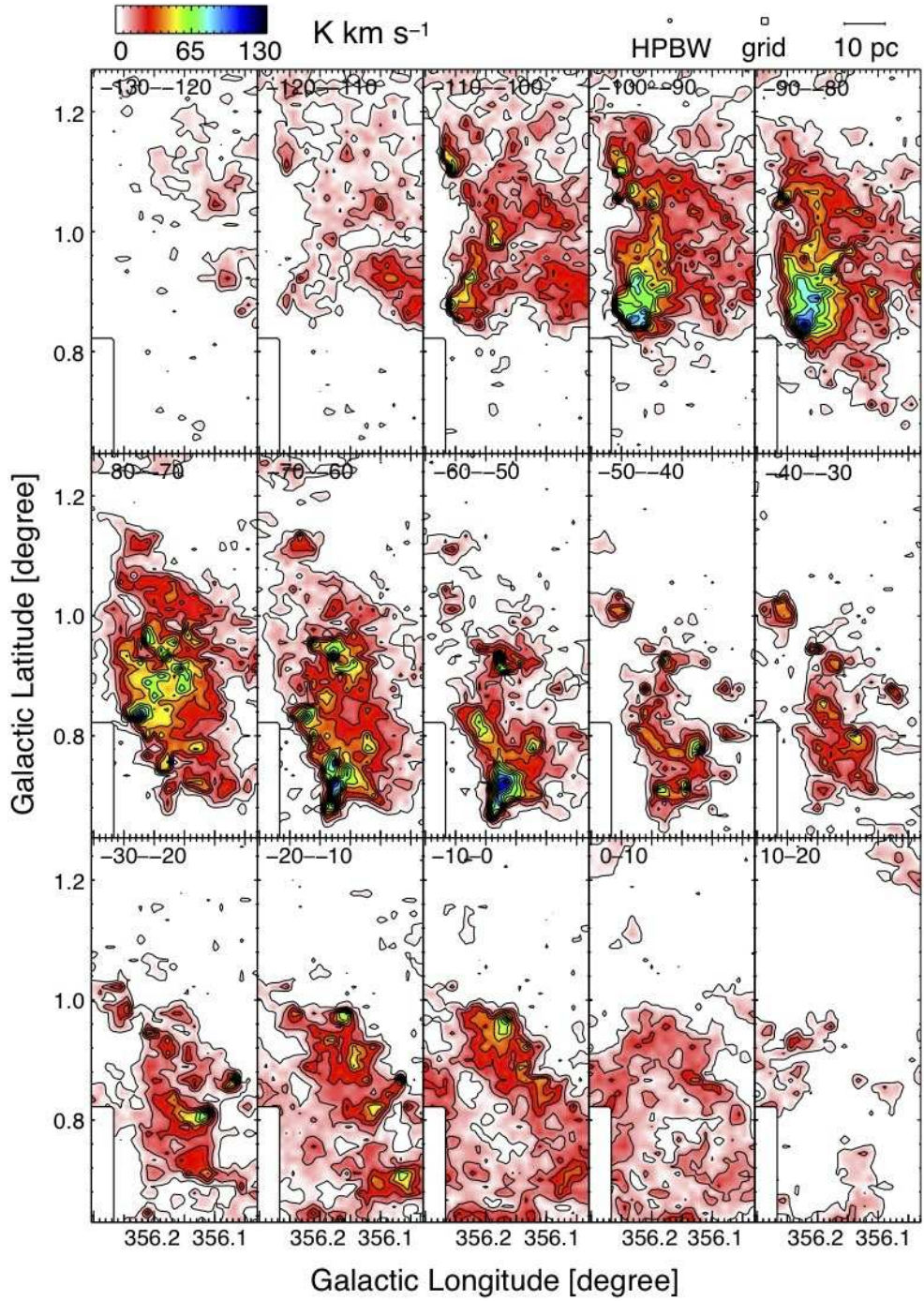}
  \end{center}
  \caption{Velocity channel maps of $^{12}$CO($J$=3--2) integrated over successive 10 km s$^{-1}$. Contours are illustrated every 10 K km s$^{-1}$ from 4 K km s$^{-1}$.}\label{fig:lb.channel.12CO3-2.2}
\end{figure}

\begin{figure}
  \begin{center}
    \FigureFile(88.38mm, 95.37mm){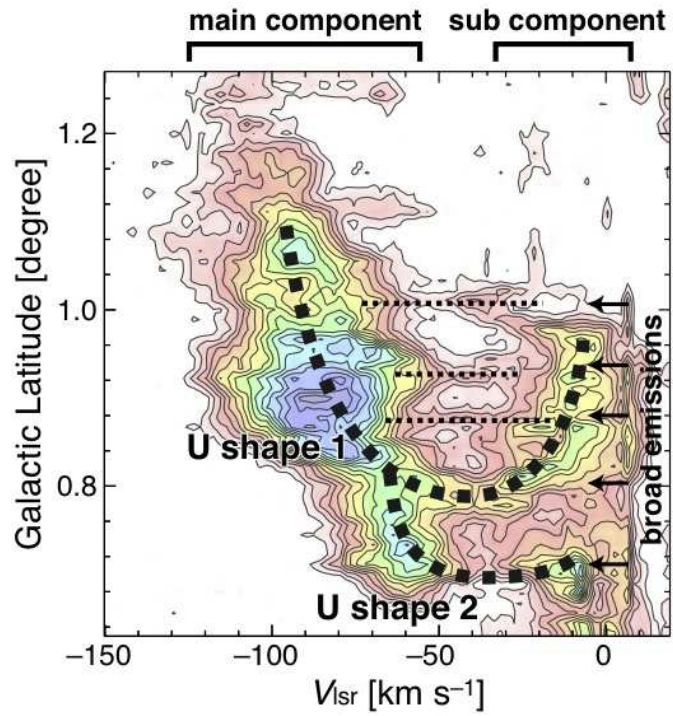}
  \end{center}
  \caption{Positions of main component, subcomponent and broad emission regions are superposed on the \atom{C}{}{12}\atom{O}{}{}($J$=3--2) figure shown in Figure \ref{fig:lvvb.3-2}.}\label{fig:lvvb.3-2.sch.eps}
\end{figure}

\begin{figure}
  \begin{center}
    \FigureFile(93.31mm, 133.59mm){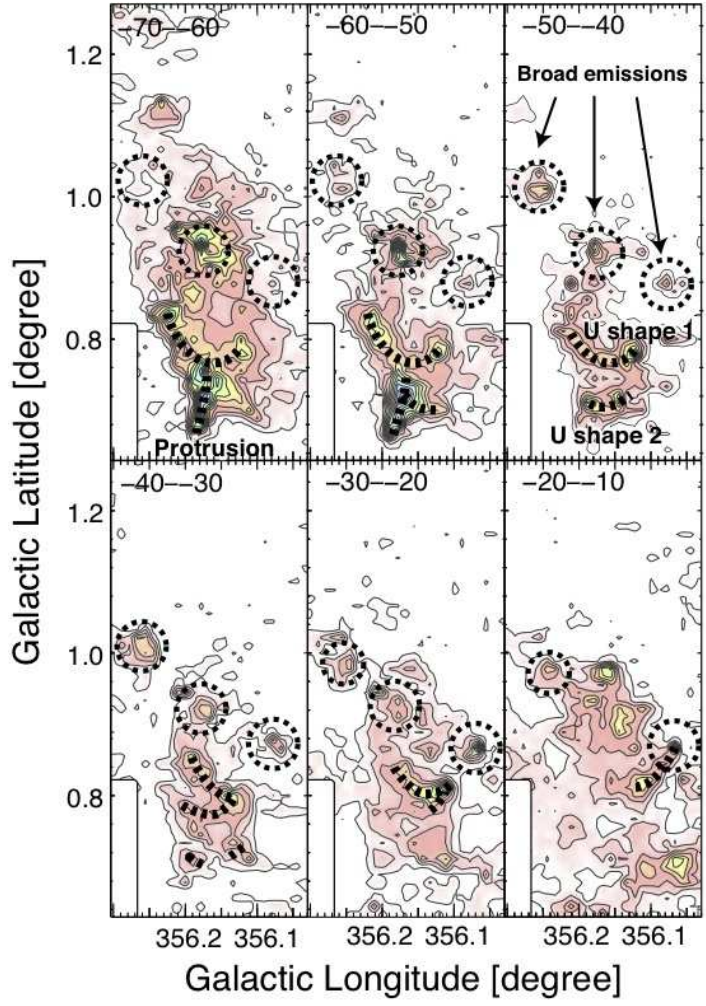}
  \end{center}
  \caption{Positions of main component, subcomponent and broad emissions are superposed on the velocity channel maps shown in Figure \ref{fig:lb.channel.12CO3-2.2}. Velocity range is from $-70$ to $-10$ km s$^{-1}$.}\label{fig:lb.channel.12CO3-2.2.sch.eps}
\end{figure}

\begin{figure}
  \begin{center}
    \FigureFile(119.19mm,64.53mm){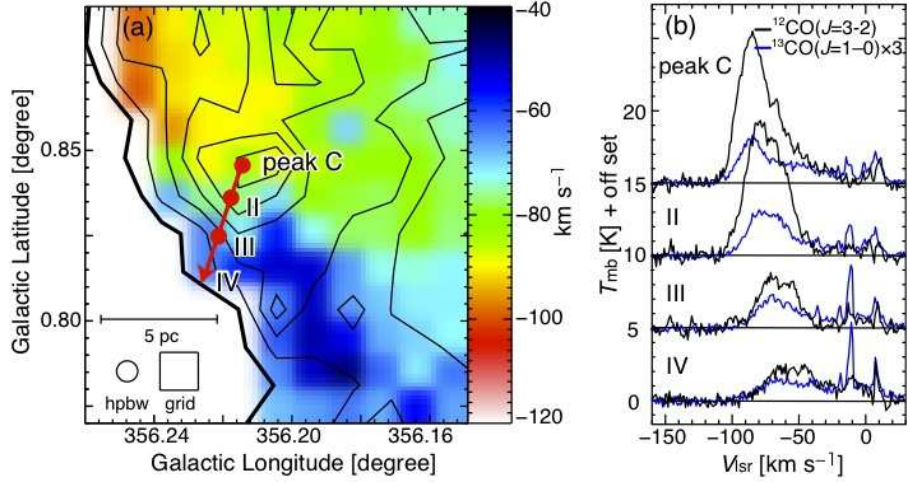}
  \end{center}
  \caption{(a) Peak velocity map around peak C estimated with $^{12}$CO($J$=3--2) is shown in the color image. Contours show the integrated intensity levels of $^{12}$CO($J$=3--2) integrated from $-140$ to $-40$ km s$^{-1}$ and are plotted every 60 K km s$^{-1}$. (b) $^{12}$CO($J$ =3--2) and $^{13}$CO($J$=1--0) spectra of the four points on the red arrow in figure (a). The order of the spectra is the direction of the red arrow, from peak C to the left-bottom of the peak C. }\label{fig:A3_3.peakV.gt75.small2.13.col}
\end{figure}

\begin{figure}
  \begin{center}
    \FigureFile(131.55mm,100.51mm){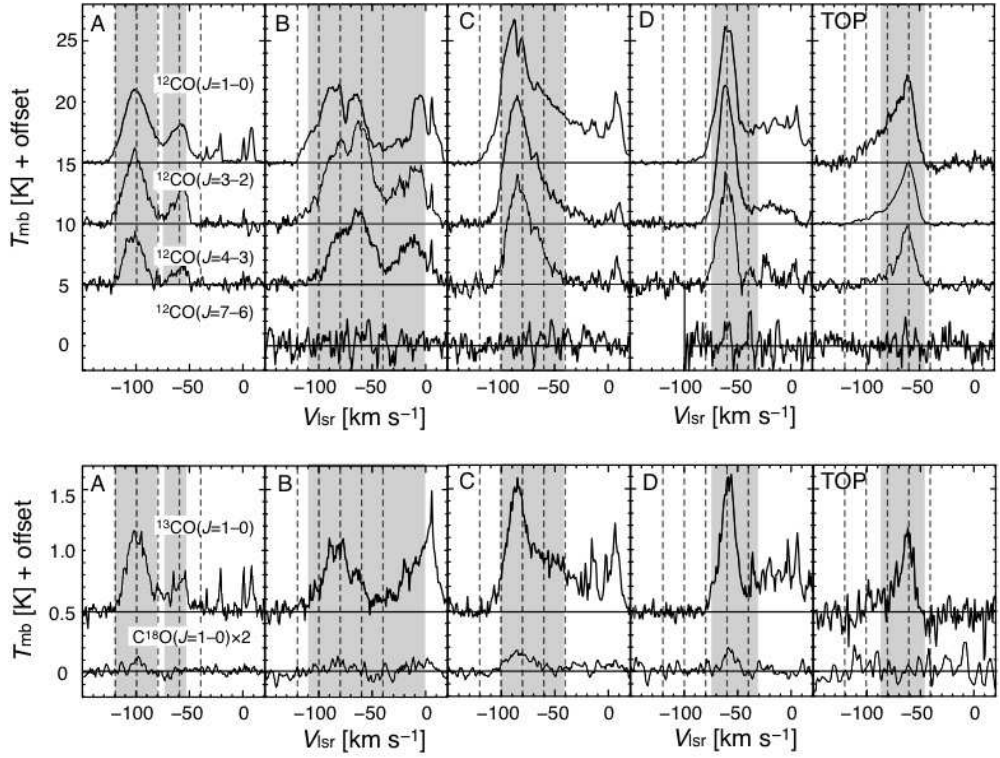}
  \end{center}
  \caption{Observed CO spectra at the peak of each of the clumps shown in Figures \ref{fig:lb.12.1-0.col} and \ref{fig:lb.13.1-0.col}. The C$^{18}$O($J$=1--0) spectra intensity scale have been multiplied by two for clearly. The vertical dashed lines are drawn with 20 km s$^{-1}$ intervals from $-120$ km s$^{-1}$. The gray scale boxes show the regions where the LVG calculations was carried out (Figure \ref{fig:specall.a-d.+LVG}).}\label{fig:specall.a-d.new}
\end{figure}

\begin{figure}
  \begin{center}
    \FigureFile(118.2mm, 85.31mm){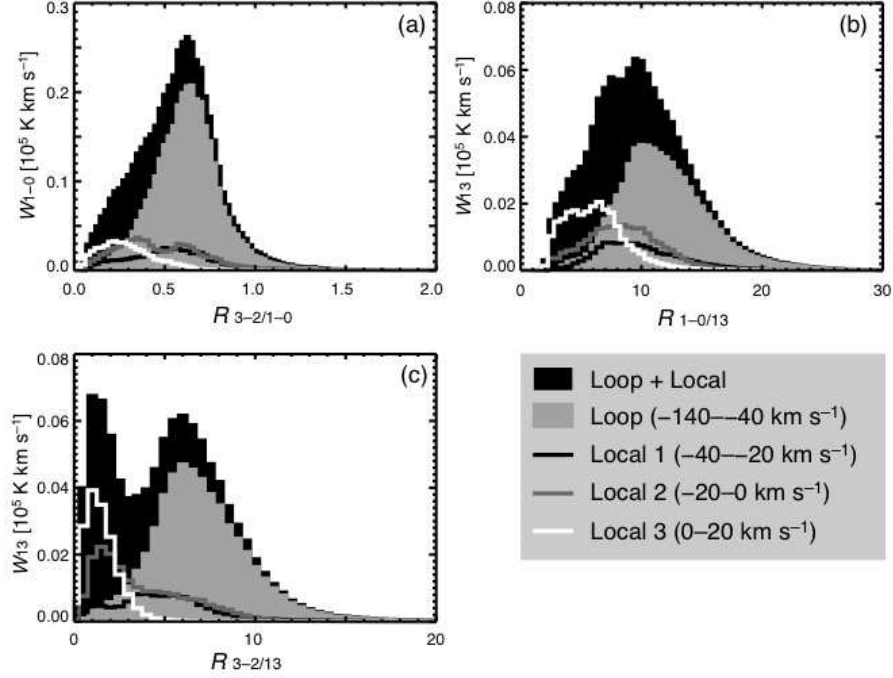}
  \end{center}
  \caption{Intensity weighted frequency distributions of the CO intensity ratio. The data were smoothed to a 60$''$ spatial resolutions with a gaussian function and smoothed to a 2 km s$^{-1}$ velocity resolution. The white lines show the contributions of the loops ($-140$ km s$^{-1}$ to $-30$ km s$^{-1}$), and the red, orange and green lines show the contributions of the local components with a interval of 20 km s$^{-1}$ from $-40$ km s$^{-1}$. (a) $R_{3-2/1-0}$ = $^{12}$CO($J$=3--2)/$^{12}$CO($J$=1--0). (b) $R_{1-0/13}$ = $^{13}$CO($J$=1--0)/$^{12}$CO($J$=1--0).  (c) $R_{3-2/13}$ = $^{13}$CO($J$=1--0)/$^{12}$CO($J$=3--2). }\label{fig:histo.ratio.3-2.1-0.13.2}
\end{figure}

\begin{figure}
  \begin{center}
    \FigureFile(69.86mm, 51.8mm){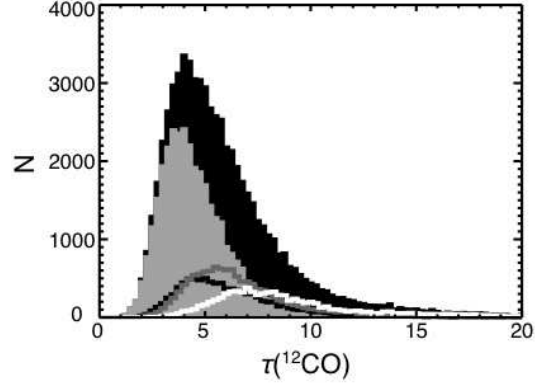}
  \end{center}
  \caption{Histogram of the $^{12}$CO($J$=1--0) optical depth in the foot point of the loops. The data are smoothed to a 5 km s$^{-1}$ resolutions. The white line shows the contributions of the loops ($-140$ km s$^{-1}$ to $-30$ km s$^{-1}$), and the red, orange and green lines show the contributions of the local components, same as in Figure \ref{fig:histo.ratio.3-2.1-0.13.2}. }\label{fig:histo.tau}
\end{figure}

\begin{figure}
  \begin{center}
    \FigureFile(124.56mm, 163.79mm){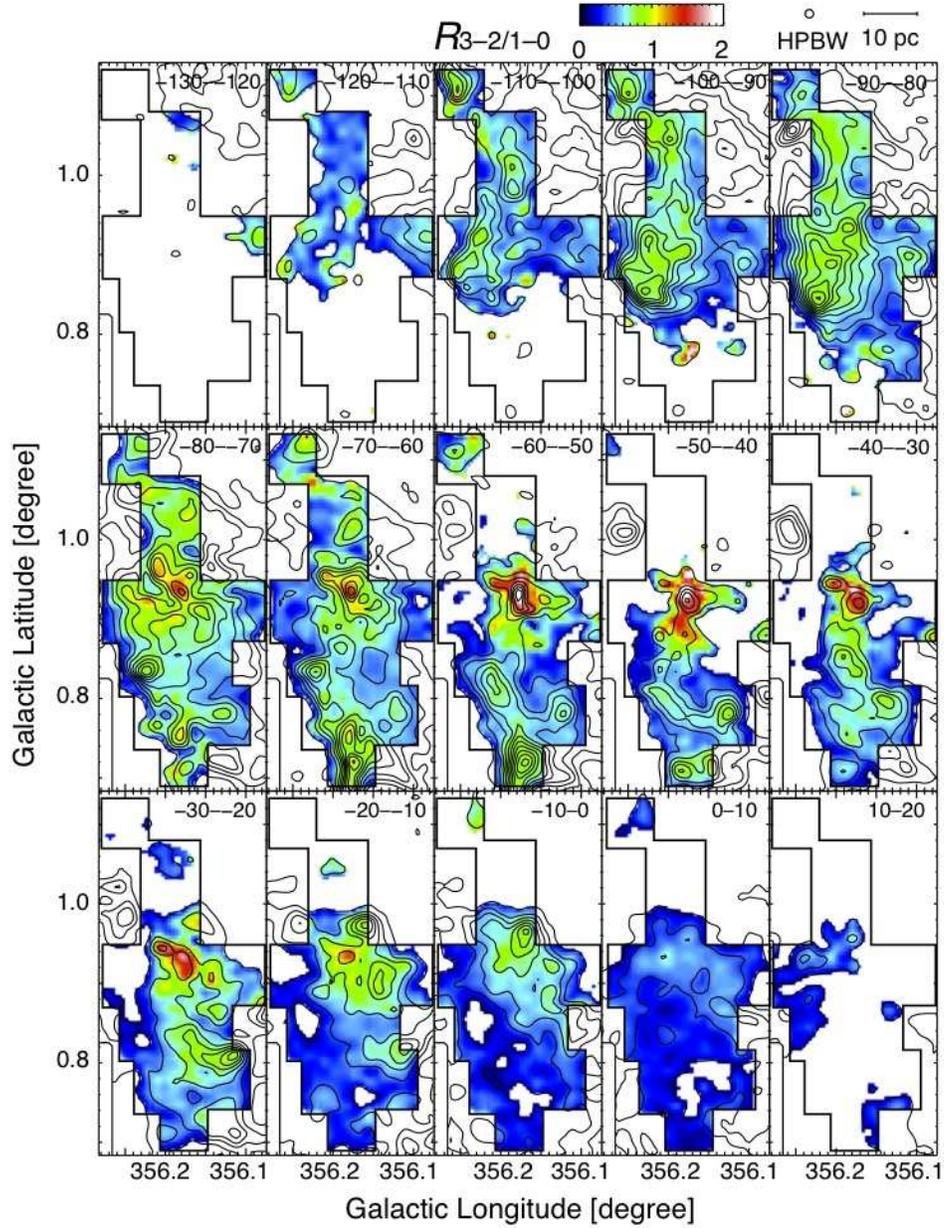}
  \end{center}
  \caption{Velocity channel maps of the $^{12}$CO($J$=3--2)/$^{12}$CO($J$=1--0) intensity ratio integrated over successive 10 km s$^{-1}$ channels. Contours are $^{12}$CO($J$=3--2) and illustrated every 10 K km s$^{-1}$ from 4 K km s$^{-1}$.}\label{fig:ratio.channel.lb.3-2.1-0.2}
\end{figure}

\begin{figure}
  \begin{center}
    \FigureFile(118.18mm, 120.36mm){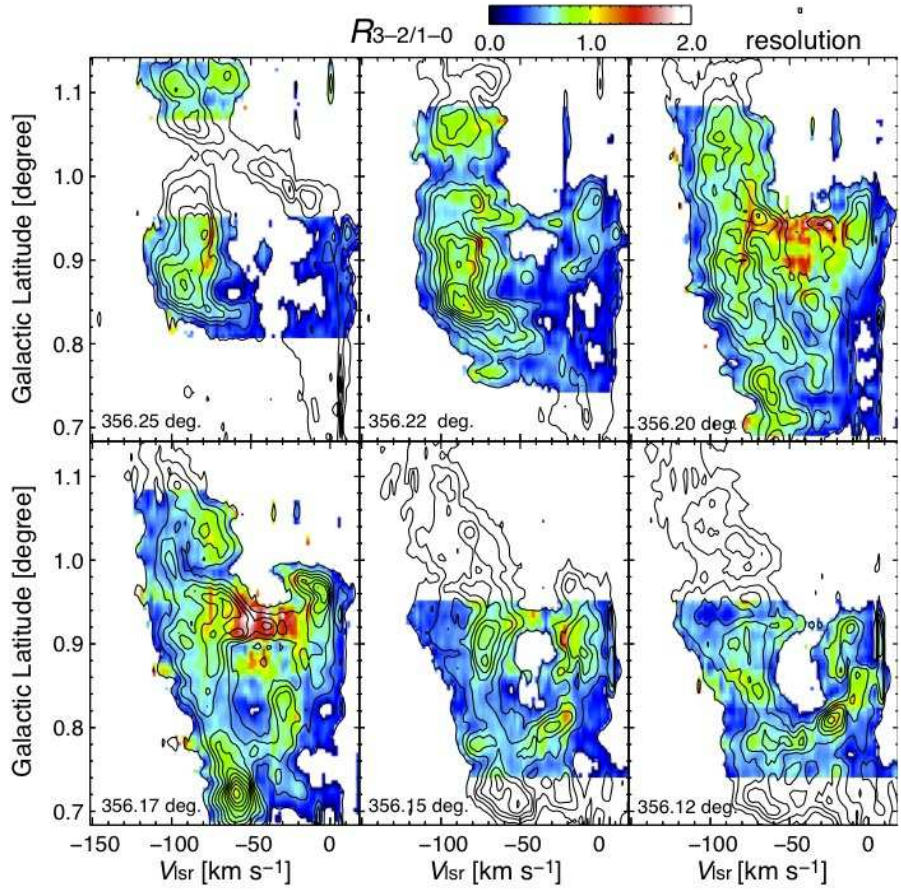}
  \end{center}
  \caption{Longitude channel maps of the $^{12}$CO($J$=3--2)/$^{12}$CO($J$=1--0) intensity ratio averaged for successive 45$''$ with an interval of 120$''$. Contours are the averaged intensity of $^{12}$CO($J$=3--2) and are plotted every 1 K from 0.5 K.}\label{fig:ratio.channel.vb.3-2.1-0.2}
\end{figure}

1) The foot points have sharp intensity gradients toward the south and east, as is consistent with the shock formation at the bottom of the foot points (Figure \ref{fig:lb.3-2.col}b). The foot points have several major peaks having 10$^4$--10$^5$ \MO gas in each. The foot points are characterized by two U shapes both in space and velocity and the protrusion (Figures 5--8). We suggest that the U shape may be formed by merging of two down flows between two loops as derived in MHD numerical simulations. 

2) Toward five selected peaks of \atom{C}{}{12}\atom{O}{}{}($J$=3--2), peaks A--D and peak top (Figures 2--4), we have carried out a multi-line LVG analysis of line radiation transfer and derived the molecular temperature and density. The four peaks in the foot point show rather high temperatures of $\sim$30--40 K and density of 10$^3$ cm$^{-3}$ to 10$^4$ cm$^{-3}$ (Figures 17 and 18). Among the four peaks, the peak toward the central region of the foot point, peak B, shows the highest temperature of $\sim100$ K or more.

3) We compared the results with calculations of C-shocks by \citet{dra1983} and find that the derived temperature and density are roughly consistent with theoretical estimates for magnetic field around 100 $\mu$G, suggesting that shock heating is a viable explanation for the high temperature over the foot point. In addition, by comparing theoretical work on the solar activity, we argue that the warmest region in the central part of the foot point may be additionally heated up either by magnetic reconnection or by upward flowing gas bounced by the narrow neck in the foot point.

\bigskip
 We thank all the members of the NANTEN2 consortium, ASTE team, and Mopra staff for the operation and persistent effors to improve the telescopes.

 The Mopra telescope is funded by the Commonwealth of Australia as a National Facility managed by CSIRO as part of the Australia Telescope. UNSW-MOPS spectrometer used was funded by the Australian Research council with the support of the Universities of New South Wales, Sydney and Macquarie, together with the CSIRO. The ASTE project is driven by Nobeyama Radio Observatory (NRO), a division of National Astronomical Observatory of Japan (NAOJ), in collaboration with University of Chile, and Japanese institutes including University of Tokyo, Nagoya University, Osaka Prefecture University, Ibaraki University, and Hokkaido Univsersity. Observations with ASTE were in part carried out remotely from Japan by using NTT's GEMnet2 and its partner R\&E (Research and Education) networks, which are based on AccessNova collaboration of University of Chile, NTT Laboratories, and NAOJ. NANTE2 project is based on a mutual agreement between Nagoya Univerisity and the University of Chile and includes member universities, Nagoya, Osaka Prefecture, Cologne, Bonn, Seoul National, Chile, New South Wales, Macquarie, Sydney and Zurich.@

 This work was financially supported in part by a Grant-in-Aid for Scientific Research (KAKENHI) from the Ministry
 of Education, Culture, Sports, Science and Technology of Japan (Nos.~15071203 and 18026004) and from JSPS (Nos.~14102003, 20244014, and 18684003). This work is also financially supported in part by core-to-core program of a Grant-in-Aid for Scientific Research from the Ministry of Education, Culture, Sports, Science and Technology of Japan (No.~17004).
 
This work was also supported by the Global COE Program of Nagoya University gQuest for Fundamental Principles in the Universe (QFPU)h from JSPS and MEXT of Japan.

LB acknowledges support from Center of Excellence in Astrophysics and Associated 
Technologies (PFB 06) and by FONDAP Center for Astrophysics 15010003.
 
\newpage


\newpage

\begin{table}
\begin{center}
\caption{Observed lines}
  \begin{tabular}{cccccc}
   \hline
   \hline
   \multirow{2}{*}{transition} & frequency & beamsize &  observing mode & noise r.m.s. & telescope\\
    & (GHz) & ($''$)  & & (K) & \\
   \hline
   
   \atom{C}{}{12}\atom{O}{}{16}($J$=1--0) & 115.27120 & 33 & On the Fly & 0.13 & Mopra\\  
   \atom{C}{}{13}\atom{O}{}{16}($J$=1--0) & 110.20137 & 33 & On the Fly & 0.06 & Mopra\\
   \atom{C}{}{12}\atom{O}{}{18}($J$=1--0) & 109.78218 & 33 & On the Fly & 0.03 & Mopra\\
   \atom{C}{}{12}\atom{O}{}{16}($J$=3--2) & 345.79599 & 22 & Position Switching & 0.33 & ASTE\\
   \atom{C}{}{12}\atom{O}{}{16}($J$=4--3) & 461.04077 & 38 & On the Fly & 0.60 & NANTEN2\\      
   \atom{C}{}{12}\atom{O}{}{16}($J$=7--6) & 806.65181 & 22 & On the Fly & 0.76 & NANTEN2\\
   
   \hline
   \multicolumn{6}{@{}l@{}}{\hbox to 0pt{\parbox{110mm}{\footnotesize
   Note. - Column (5): Typical noise r.m.s. per channel. The channel width of all the data is 0.86 km s$^{-1}$.
   }\hss}} \label{table:obsline}
   \end{tabular}
\end{center}
\end{table}

\begin{table}
\begin{center}
\caption{Observed areas of position-switch observaions}
   \begin{tabular}{cccccc}
   \hline
   \hline

	\multirow{2}{*}{transition} & \multirow{2}{*}{coord.} &  \multirow{2}{*}{observed area} & grid 
	      & \multirow{2}{*}{$N_{\mathrm{point}}$} &  \multirow{2}{*}{region} \\
	& & & ($''$) && \\
	\hline
	
	\multirow{3}{*}{ \atom{C}{}{12}\atom{O}{}{}($J$=3--2) } & \multirow{2}{*}{galactic} 
	& (356.033$^\circ$, 0.633$^\circ$)--(356.267$^\circ$, 0.811$^\circ$) & \multirow{2}{*}{40} & \multirow{2}{*}{1399} & \multirow{2}{*}{foot point}\\
	& & (356.267$^\circ$, 0.822$^\circ$)--(356.300$^\circ$, 1.267$^\circ$) & &  \\
	& galactic & (356.462$^\circ$, 1.191$^\circ$)--(356.591$^\circ$, 1.492) & 40 & 2336 & top \\

   \hline
   
    \multicolumn{5}{@{}l@{}}{\hbox to 0pt{\parbox{140mm}{\footnotesize
   	Note. - Column (2) : Coordinate system of observations. Column (4) : Grid intervals of observations. Column (5) : Number of total observed points.}\hss}}
   \label{tab:obspsw}
   \end{tabular}

\end{center}
\end{table}

\begin{table}
\begin{center}
\caption{Observed areas of On-the-fly observations}
   \begin{tabular}{ccccccc}
   \hline
   \hline
   \multirow{2}{*}{transition} & \multirow{2}{*}{coord.} &  \multirow{2}{*}{reference position} & map size & grid  
   & \multirow{2}{*}{$N_{\mathrm{map}}$} & \multirow{2}{*}{region}\\
    &  &  & ($'$) & ($''$) && \\
   \hline
   
  \multirow{2}{*}{ \shortstack{ \atom{C}{}{12}\atom{O}{}{}, \atom{C}{}{13}\atom{O}{}{} \\ and \atom{C}{}{}\atom{O}{}{18}($J$=1--0) } }
	& \multirow{2}{*}{galactic}  & 356.222$^{\circ}$, 0.845$^{\circ}$ & 3 - 4 & 15 & 11 & foot point\\
      && 356.556$^{\circ}$, 1.333$^{\circ}$ & 2 & 15 & 1 & top\\

  \hline
  \multirow{5}{*}{ \shortstack{ \atom{C}{}{12}\atom{O}{}{}($J$=4--3) \\ \atom{C}{}{12}\atom{O}{}{}($J$=7--6) }}
  	& \multirow{5}{*}{Equatorial(J2000.0)} 
	   & \timeform{17h31m54.58s}, \timeform{-31D08'13.76''} & 2 & 15 & 1 & top\\
	& & \timeform{17h32m03.68s}, \timeform{-31D30'55.44''} & 2 & 15 & 1 & peak A\\
	& & \timeform{17h32m30.36s}, \timeform{-31D40'04.02''} & 2 & 15 & 1 & peak B\\
	& & \timeform{17h32m58.84s}, \timeform{-31D40'59.49''} & 2 &10 & 1 & peak C\\   
	& & \timeform{17h33m22.07s}, \timeform{-31D47'51.85''} & 2 & 15 & 1 & peak D\\   
   
   \hline

 \multicolumn{6}{@{}l@{}}{\hbox to 0pt{\parbox{140mm}{\footnotesize
   	Note. - Column (2) : Coordinate system of observations. Column (4) :  Size of the unit square of OTF scan. Column (5) : Grid intervals of the out put data. Column (6) : Number of maps that were observed with each reference position.}\hss}}
	\label{tab:obsotf}
 
  \end{tabular}  
\end{center}
\end{table}

\begin{table}
\begin{center}
\caption{10 km s$^{-1}$ average intensity of the peaks.}
  \begin{tabular}{crrrrrr}
   \hline
   \hline
           \multirow{2}{*}{Peak} & $V_{\mathrm{lsr}}$ & \multicolumn{5}{c}{10 km s$^{-1}$ average intensity (K)}\\
           \cline{3-7}
           & (km s$^{-1}$) & $^{12}$CO(1--0) & $^{12}$CO(3--2) & $^{12}$CO(4--3) & $^{13}$CO(1--0) & C$^{18}$O(1--0)\\
                   
            \hline \multirow{6}{*}{A} 
            &	$-110$	& 3.31	& 3.31	& 2.29	&0.21	&---\\
            &	$-100$ 	& 5.26	&5.49	&4.00	&0.63	&0.05\\
            &	$-90$	& 4.27	&3.41	&2.24	&0.51	&---\\
            &	$-80$	& 2.40	&1.44	&0.71	&0.20	&---\\
            &	$-68$	& 2.06	&1.23	&0.61	&0.14	&---\\
            &	$-58$	& 2.43	&1.88	&0.90	&0.21	&---\\
            
            \hline \multirow{11}{*}{B}  
            & $-105$ 	& 2.39	&1.39	&0.54	&0.15	&---\\
            & $-95$		& 4.75	&3.77	&1.31	&0.36	&---\\
            & $-85$ 		& 6.56	&6.00	&3.10	&0.52	&--- \\
            & $-75$ 		& 4.52	&5.76	&3.54	&0.40	&---\\
            & $-65$ 		& 5.24	&7.10	&4.81	&0.33	&---\\
            & $-55$ 		& 2.84	&5.48	&3.62	&0.15	&---\\
            & $-45$ 		& 0.87	&1.94	&1.45	&0.08	&---\\
            & $-35$ 		& 0.82	&1.51	&0.67	&0.14	&---\\
            & $-25$ 		& 1.62	&2.50	&1.85	&0.20	&---\\
            & $-15$ 		& 2.58	&3.42	&2.44	&0.39	&---\\
            & $-5$ 		& 5.07	&4.30	&2.85	&0.54	&---\\
            
           \hline \multirow{6}{*}{C} 
           &  $-95$ 		& 8.90	&6.21	&4.41	&0.53	&0.04\\
           & $-85$		& 10.33	&9.50	&7.89	&0.92	&0.08\\
           & $-75$ 		& 8.22	&7.54	&6.05	&0.69	&0.06\\
           & $-65$ 		& 6.75	&5.62	&4.08	&0.54	&0.04\\
           & $-55$ 		& 5.34	&3.19	&1.87	&0.42	&--- \\
           & $-45$ 		&  4.45	&1.84	&0.94	&0.38	&---\\
                      
           \hline \multirow{4}{*}{D} 
           &  $-68$		& 6.61	&5.83	&4.02	&0.50	&---\\
           & $-58$ 		& 10.95	&10.65	&8.10	&0.99	&0.10\\
           &  $-48$		& 5.42	&4.75	&3.19	&0.47	&--- \\
           &  $-38$		& 2.77	&1.02	&0.77	&0.29	&--- \\
           
           \hline \multirow{4}{*}{TOP} 
           & $-80$ 		& 3.35	&1.36	&1.00	&0.24	&---\\
           & $-70$ 		& 4.83	&2.84	&2.28	&0.50	&--- \\
           & $-60$ 		& 6.23	&4.46	&4.06	&0.69	&--- \\
           & $-50$ 		& 2.09	&1.66	&1.32	&0.13	&---\\          
           
           \hline
   \multicolumn{6}{@{}l@{}}{\hbox to 0pt{\parbox{120mm}{\footnotesize
     \par\noindent
 Note.  "---" stands for non-detection above 3$\sigma$.
	Column (2) indicates Center velocity for calculations.
   }\hss}} \label{tab:ratio}
   
   \end{tabular}
\end{center}
\end{table}

\begin{table}
\begin{center}
\caption{Physical properties of the clumps.}
  \begin{tabular}{ccccccc}
   \hline
   \hline
          $l$ & $b$ & $\Delta$V & $r$ & M$_{\mathrm{vir}}$ & M$_{\atom{C}{}{}\atom{O}{}{}}$ & \multirow{2}{*}{region} \\
          ($^\circ$) &  ($^\circ$) & (km s$^{-1}$) & (pc) & ($\times$ 10$^4 \MO$) & ($\times$ 10$^4 \MO$) &\\
   \hline
     
     356.124	&	0.780	&	26.0	&	4.6	&	65.0	&	1.2\footnotemark[$\dagger$] /\ 2.5\footnotemark[$\ddagger$] 	&	\\
     356.149	&	0.913	&	30.4&	2.9	&	56.0	&	0.7\footnotemark[$\dagger$] /\ 1.4\footnotemark[$\ddagger$] 	&	peak B\\
     356.174	&	0.708	&	19.2	&	2.8	&	21.6&	0.6\footnotemark[$\dagger$] /\ 1.2\footnotemark[$\ddagger$] 	&	peak D\\
     356.187	&	0.988	&	33.1	&	2.4	&	55.0	&	0.4\footnotemark[$\dagger$] /\ 0.8\footnotemark[$\ddagger$] 	&	\\
     356.220	&	0.838	&	36.0	&	5.9	&	159.8&	3.7\footnotemark[$\dagger$] /\ 7.4\footnotemark[$\ddagger$] 	&	peak C\\
     356.254	&	1.113	&	38.2	&	3.9	&	118.9&	1.5\footnotemark[$\dagger$] /\ 3.0\footnotemark[$\ddagger$] 	&	peak A\\

   \hline
   
   \multicolumn{5}{@{}l@{}}{\hbox to 0pt{\parbox{115mm}{\footnotesize
   \footnotemark[$\dagger$] [H$_2$]/[$^{13}$CO]$=5\times10^5$\citep{dic1978} was used for estimation.\\
   \footnotemark[$\ddagger$] [H$_2$]/[$^{13}$CO]$=10^6$\citep{lis1989} was used for estimation.\\
   	Note. -  Column (1, 2) : Peak position of the clump. Column (3) : Intensity weighted velocity dispersion (FWHM). Column (4) : Radius of the clumps. Column (5) : Clump mass derived with assumption of Virial equilibrium. $M_{\mathrm{vir}} = 209 r (\Delta V)^2$. Column (6) : Clump mass derived by using $^{13}$CO($J$=1--0) with LTE assumption. }\hss}}\label{table:clumpphys}
   
   \end{tabular}
\end{center}
\end{table}

\begin{table}
\begin{center}
\caption{LVG results for X(CO)/(dv/dr) = 1.1 $\times$ 10$^{-5}$.}
  \begin{tabular}{ccrccrcc}
   \hline
   \hline
           \multirow{2}{*}{Peak}  & V$_{\mathrm{lsr}}$
           & \multicolumn{2}{c}{n(\atom{H}{}{}$_2$) (cm$^{-3}$)}
           && \multicolumn{2}{c}{$T_{\mathrm{k}}$ (K)}  & \multirow{2}{*}{min. $\chi^2$ } \\
           \cline{3-4} \cline{6-7}
           & (km s$^{-1}$)& $\chi^2<25$ & min. $\chi^2$ 
           && $\chi^2<25$ & min. $\chi^2$ \\
 
   \hline
  
  \multirow{5}{*}{A} &
  $-110$	&	$10^{3.1}$--$10^{4.2}$	&	$10^{3.6}$	&&	38--128 		&53		&2.88\\
  &$-100$&	$10^{3.8}$--$10^{4.0}$	&	$10^{3.9}$	&&	35--41		&38		&22.84\\
  &$-90$	&	$10^{3.5}$--$10^{3.9}$	&	$10^{3.6}$	&&	24--36		&28		&15.77\\ 
  &$-80$	&	$10^{2.9}$--$10^{3.6}$	&	$10^{3.3}$	&&	19--65		&27		&3.24\\
  &$-68$	&	$10^{2.4}$--$10^{3.7}$	&	$10^{3.2}$	&&	$>$ 20	 	&32		&3.35\\
  &$-58$	&	$10^{3.0}$--$10^{3.8}$	&	$10^{3.4}$	&&	22--61		&30		&4.45\\
  \hline
  
  \multirow{11}{*}{B} &
  $-105$	&	$10^{2.4}$--$10^{3.5}$	&	$10^{3.1}$	&&	$>$ 20		&33		&3.34\\
  &$-95$	&	$10^{3.1}$--$10^{3.4}$	&	$10^{3.3}$	&&	27--40		&32		&17.16\\
  &$-85$	&	$10^{3.4}$--$10^{3.5}$	&	$10^{3.4}$	&&	37--38		&37		&24.61\\ 
  &$-75$	&	---					&	---			&&	---			&---		&30.52\\
  &$-65$	&	$10^{3.5}$--$10^{3.8}$	&	$10^{3.7}$	&&	57--76		&65		&18.85\\
  &$-55$	&	$10^{3.3}$--$10^{3.7}$	&	$10^{3.5}$	&&	$>$ 97		&128	&22.55\\
  &$-45$	&	$10^{3.2}$--$10^{4.3}$	&	$10^{3.7}$	&&	$>$ 60		&98		&19.68\\ 
  &$-35$	&	$10^{4.0}$--$10^{4.4}$	&	$10^{4.2}$	&&	44--59		&48		&24.76\\
  &$-25$	&	$10^{3.8}$--$10^{4.7}$	&	$10^{4.0}$	&&	41--91		&51		&20.75\\
  &$-15$	&	---					&	---			&&	---			&---		&30.36\\
  &$-5$	&	$10^{3.5}$--$10^{3.8}$	&	$10^{3.6}$	&&	28 - 37		&31		&19.24\\ 
  \hline
   
  \multirow{6}{*}{C} &
  $-95$	&	$10^{3.2}$--$10^{3.5}$	&	$10^{3.3}$	&&	41--56		&47		&13.13\\
  &$-85$	&	$10^{3.8}$--$10^{3.8}$	&	$10^{3.8}$	&&	42--43		&42		&24.59\\
  &$-75$	&	$10^{3.5}$--$10^{3.9}$	&	$10^{3.8}$	&&	38--50		&44		&11.78\\ 
  &$-65$	&	$10^{3.4}$--$10^{3.8}$	&	$10^{3.6}$	&&	34--48		&40		&11.76\\
  &$-55$	&	$10^{3.2}$--$10^{3.5}$	&	$10^{3.3}$	&&	26--39		&31		&10.68\\
  &$-45$	&	$10^{3.1}$--$10^{3.3}$	&	$10^{3.2}$	&&	19--28		&22		&13.85\\
  \hline
  
  \multirow{4}{*}{D} &
  $-68$	&	$10^{3.4}$--$10^{3.8}$	&	$10^{3.5}$	&&	37--50		&42		&13.44\\
  &$-58$	&	$10^{3.7}$--$10^{3.8}$	&	$10^{3.8}$	&&	39--43		&41		&22.30\\
  &$-48$	&	$10^{3.4}$--$10^{3.9}$	&	$10^{3.6}$	&&	32--46		&37		&13.02\\ 
  &$-38$	&	$10^{3.1}$--$10^{3.4}$	&	$10^{3.3}$	&&	17--24		&20		&17.55\\
  \hline

  \multirow{3}{*}{TOP} &
  $-80$	&	$10^{2.2}$--$10^{3.4}$	&	$10^{3.0}$	&&	$>$ 18		&37		&12.88\\
  &$-70$	&	$10^{3.2}$--$10^{3.8}$	&	$10^{3.8}$	&&	20--47		&26		&10.00\\
  &$-60$	&	$10^{3.4}$--$10^{4.1}$	&	$10^{3.8}$	&&	24--46		&32		&5.75\\
  &$-50$	&	$> 10^{2.6}$			&	$10^{3.6}$	&& $>$ 22		&42		&4.67\\ 
  \hline
    \multicolumn{8}{@{}l@{}}{\hbox to 0pt{\parbox{120mm}{\footnotesize
   \par\noindent
 	Note. - "---" stands for no-solution above 5\% confidence level of $\chi^2$ distribution.
	Column (3): Number density, $n(\atom{H}{}{}_2)$ cm$^{-3}$, for that $\chi^2$ is less than 25. 
	(4) : Number density, $n(\atom{H}{}{}_2)$ cm$^{-3}$, at the point that minimum $\chi^2$ is found. 
	(5) : Kinetic temperature, $T_{\mathrm{k}}$ K, for that $\chi^2$ is less than 25. 
	(6) : Kinetic temperature, $T_{\mathrm{k}}$ K, at the point that minimum $\chi^2$ is found. 
	(7) : minimum $\chi^2$
	}\hss}} \label{tab:lvg}
   \end{tabular}
\end{center}
\end{table}

\begin{figure}
  \begin{center}
    \FigureFile(130.76mm,85.96mm){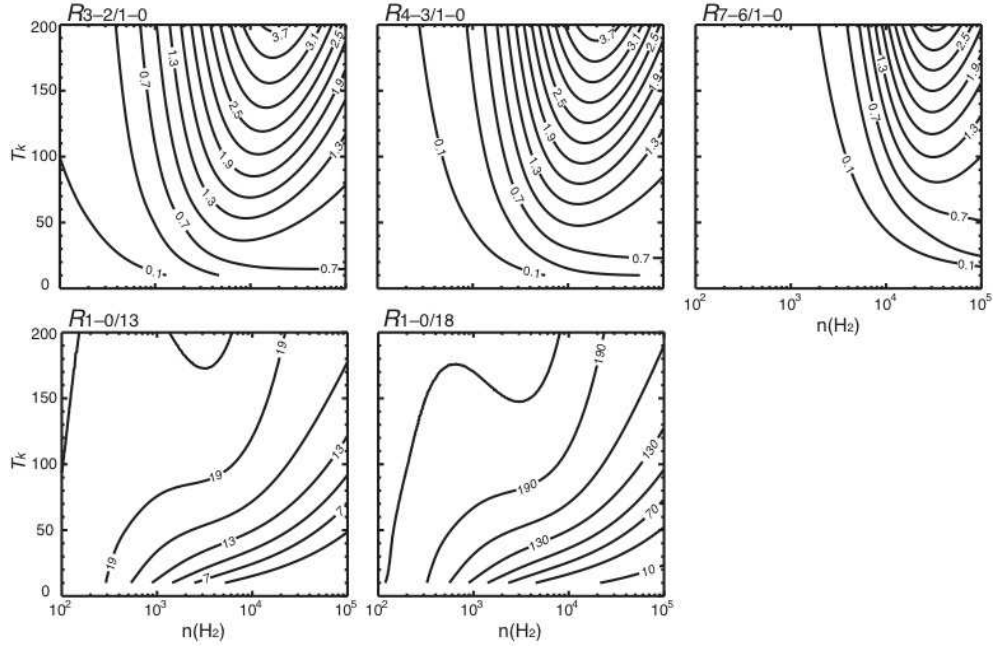}
  \end{center}
  \caption{Intensity ratio distributions in the density-temperature space calculated from the LVG model for X(CO)/(dv/dr) of 1.1 $\times$ 10$^{-5}$. $R_{3-2/1-0}$,  $R_{4-3/1-0}$, $R_{7-6/1-0}$, $R_{1-0/13}$ and $R_{1-0/18}$ stand for the intensity ratio of $^{12}$CO($J$=3--2)/$^{12}$CO($J$=1--0), $^{12}$CO($J$=4--3)/$^{12}$CO($J$=1--0), $^{12}$CO($J$=7--6)/$^{12}$CO($J$=1--0), $^{12}$CO($J$=1--0)/$^{13}$CO($J$=1--0) and $^{12}$CO($J$=1--0)/C$^{18}$O($J$=1--0), respectively.}\label{fig:lvg.ratio_plot}
\end{figure}

\begin{figure}
  \renewcommand{\thefigure}{17a}
  \begin{center}
  \FigureFile(127.28mm,90.75mm){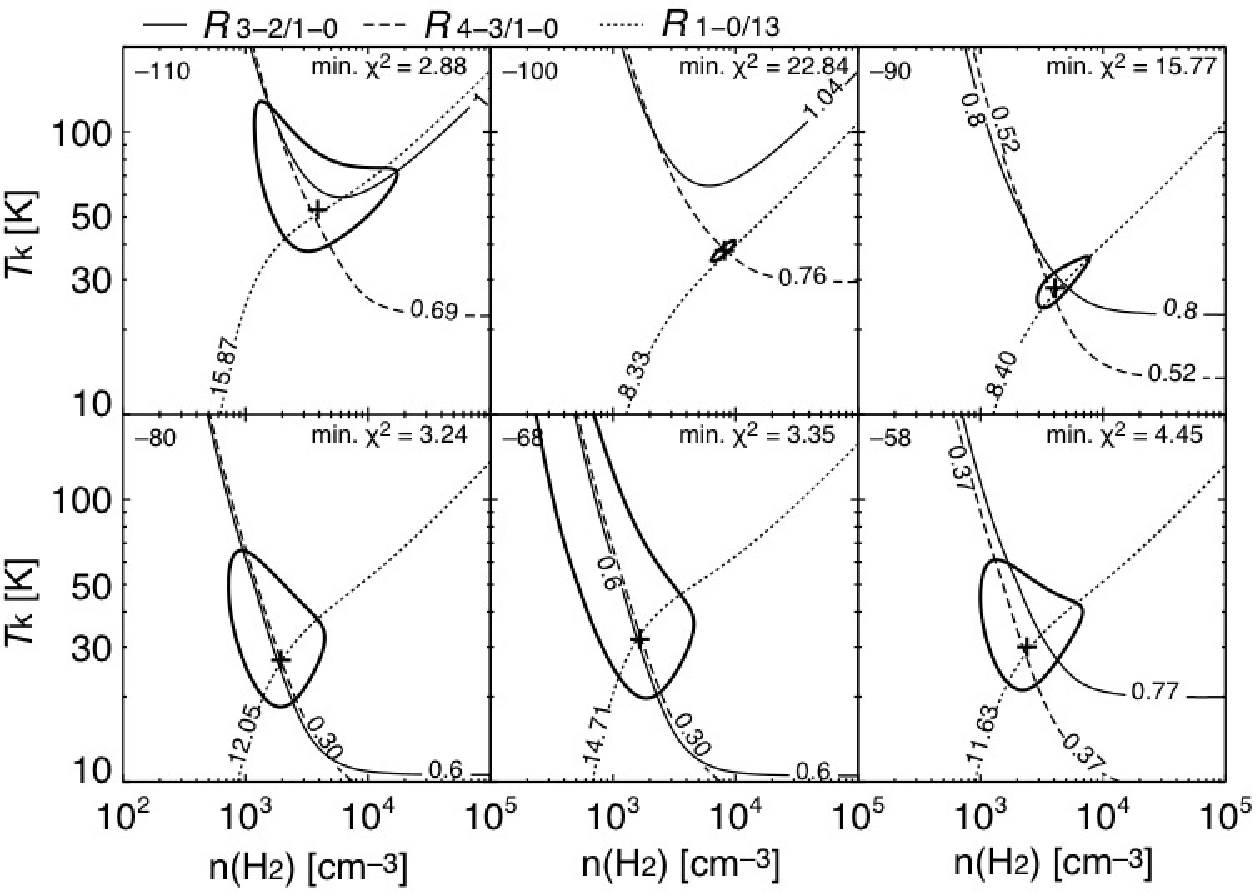}
  \end{center}
  \contcaption{LVG results of peak A. Values shown at left-top of the figures show the center velocity of the calculated velocity range. Crosses denote the lowest point of chi-square $\chi^2$. Contours surrounding the cross indicate $\chi^2=2$5.0 which correspond to 5\% confidence level of $\chi^2$ distribution with 15 degree of freedom. Thin lines show the typical intensity ratios; $R_{3-2/1-0}$ (solid lines), $R_{4-3/1-0}$(dashed lines) and $R_{1-0/13}$(dotted lines).}\label{fig:LVGall.peakA}
\end{figure}

\begin{figure}
  \renewcommand{\thefigure}{17b}
  \begin{center}
  \FigureFile(126.53mm,165.08mm){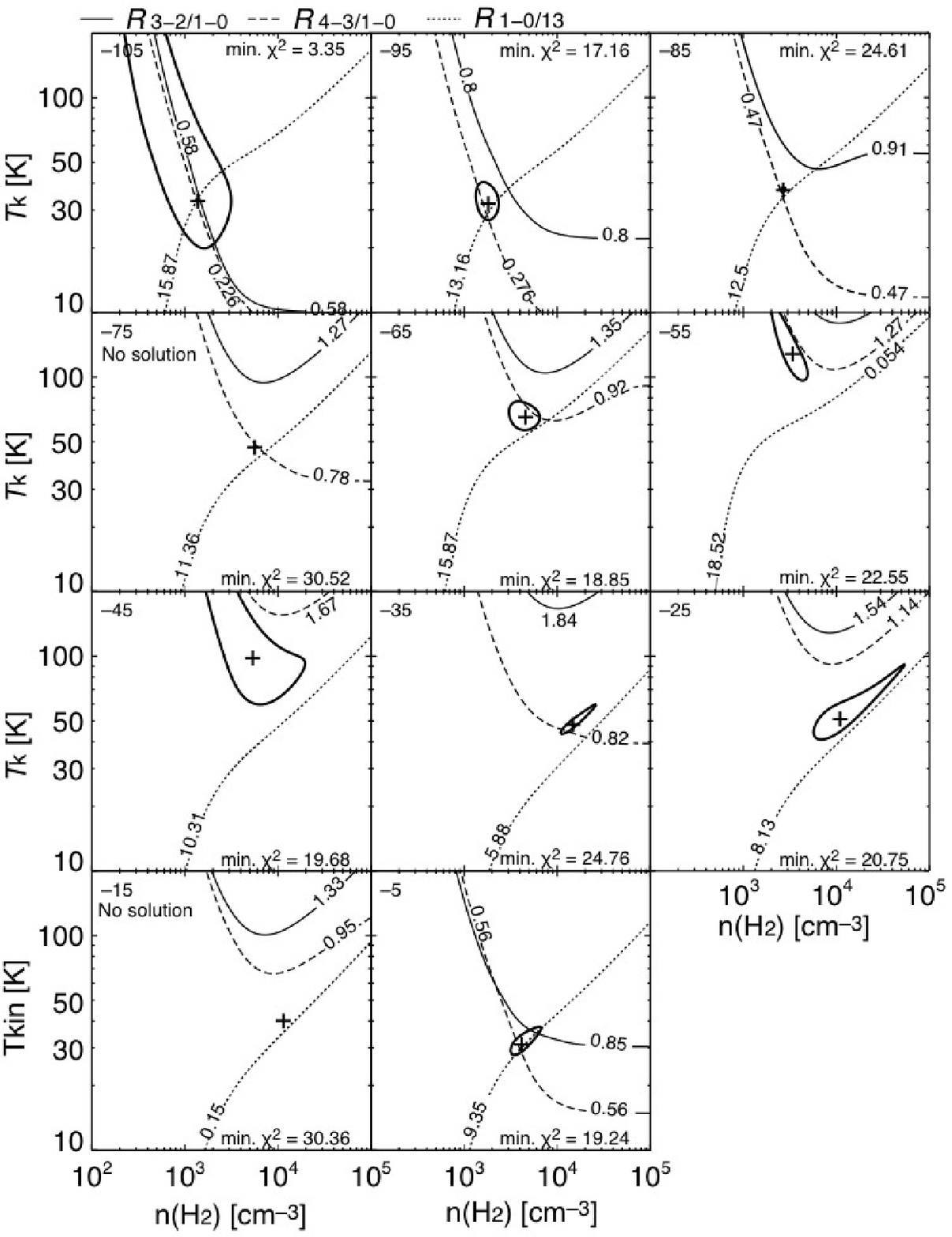}
  \end{center}
  \contcaption{LVG results of peak B. Details are the same in Figure 17a.}\label{fig:LVGall.peakB}
\end{figure}

\begin{figure}
  \renewcommand{\thefigure}{17c}
  \begin{center}
  \FigureFile(127.05mm,91.33mm){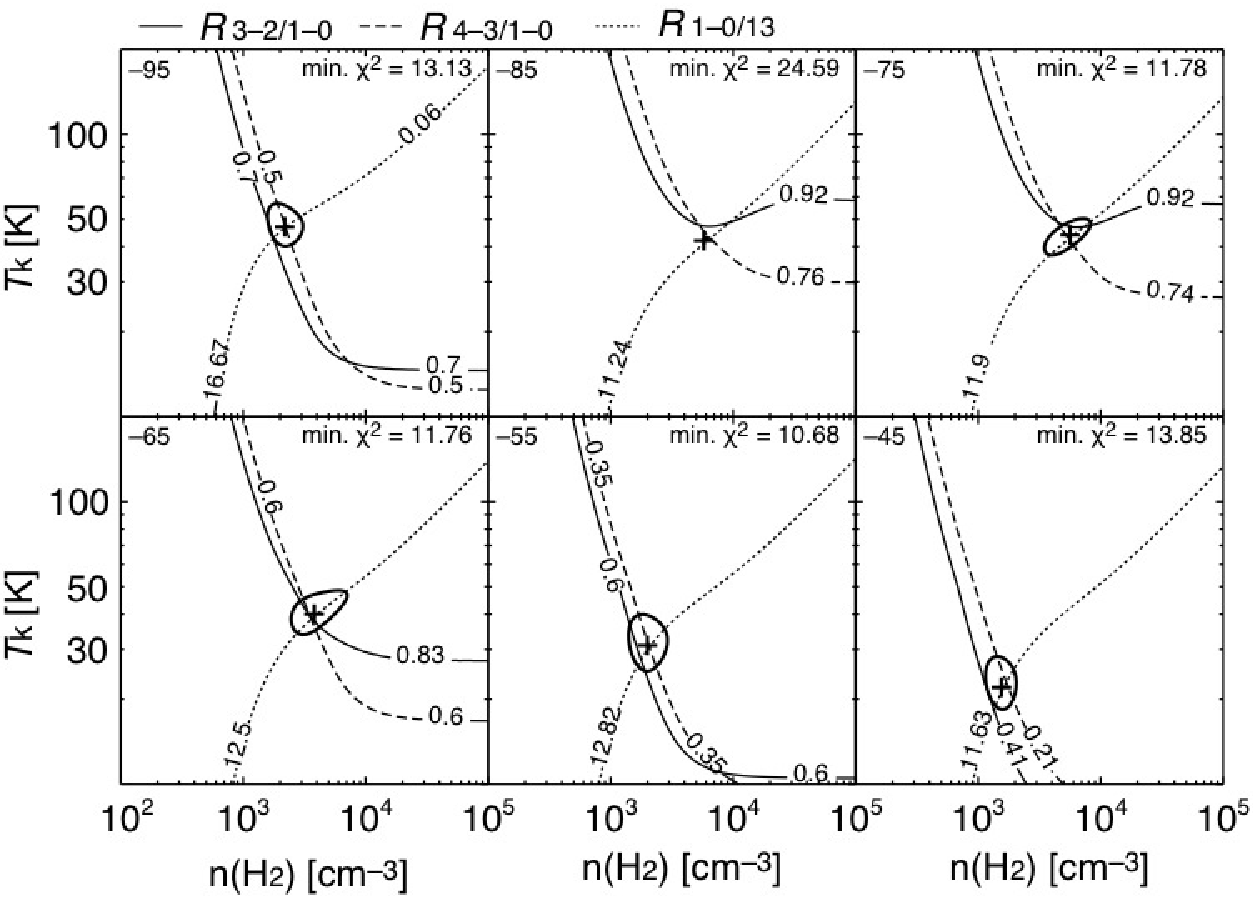}
  \end{center}
  \contcaption{LVG results of peak C. Details are the same in Figure 17a.}\label{fig:LVGall.peakC}
\end{figure}

\begin{figure}
  \renewcommand{\thefigure}{17d}
  \begin{center}
  \FigureFile(90mm,90.27mm){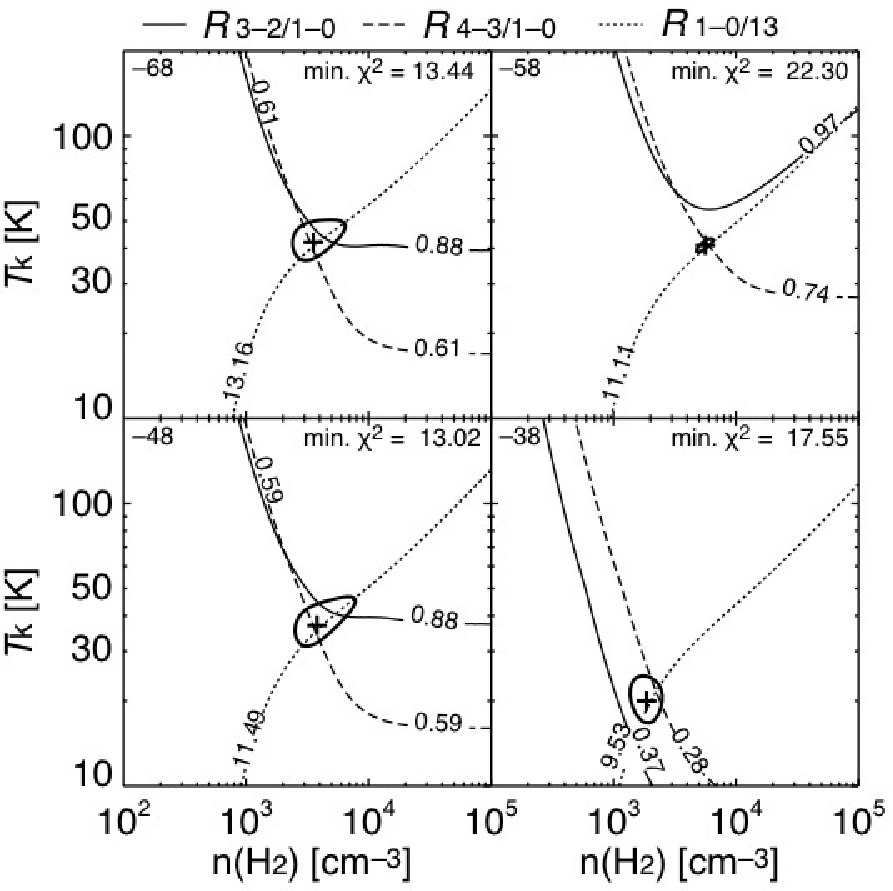}
  \end{center}
  \contcaption{LVG results of peak D. Details are the same in Figure 17a.}\label{fig:LVGall.peakD}
\end{figure}

\begin{figure}
  \renewcommand{\thefigure}{17e}
  \begin{center}
  \FigureFile(89.67mm,89.56mm){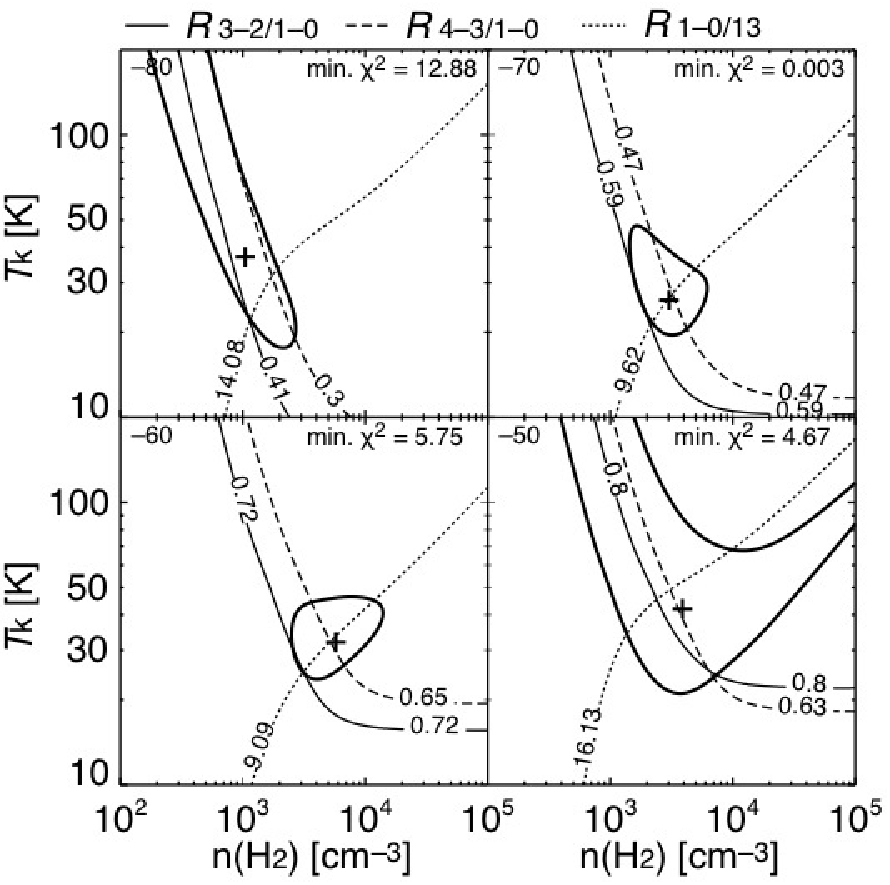}
  \end{center}
  \contcaption{LVG results of peak top. Details are the same in Figure 17a.}\label{fig:LVGall.peakTOP}
\end{figure}

\begin{figure}
  \renewcommand{\thefigure}{18}
  \begin{center}
    \FigureFile(131.57mm,97.28mm){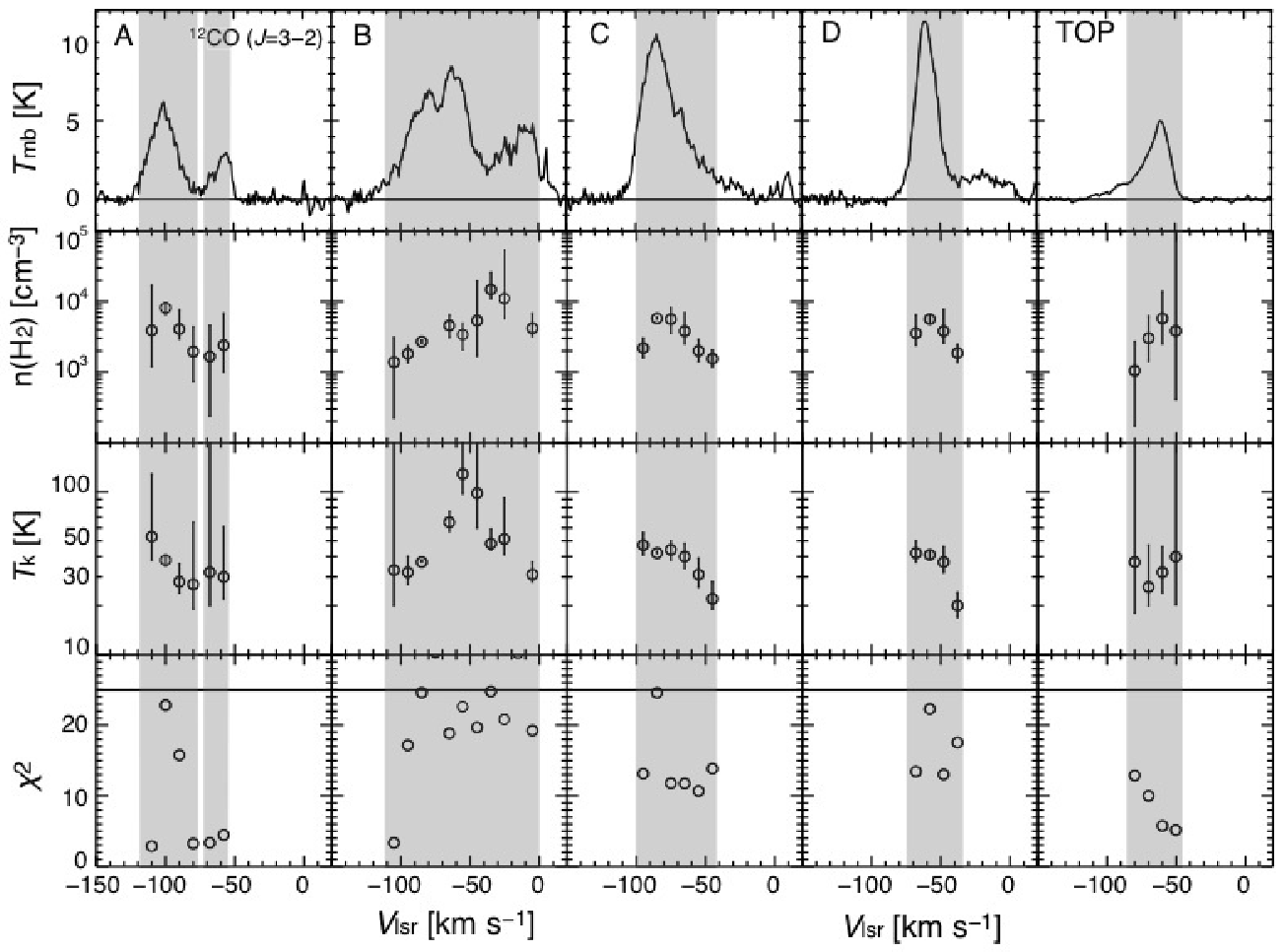}
  \end{center}
  \caption{LVG results for X(CO)/(dv/dr) of 1.1 $\times 10^{-5}$ for the five spectra at the peaks. The horizontal axis of all figures is LSR velocity. The top row shows the \atom{C}{}{12}\atom{O}{}{}($J$=3--2) spectra at the peaks A-D, and the peak in the loop top. The second and third rows show the number density, $n$(H$_2$) cm$^{-3}$ and the kinetic temperature, $T_\mathrm{k}$ K, respectively. Circles show the lowest point of $\chi^2$, and the error range is defined as a 5\% confidence level of $\chi^2$ distribution with 15 degree of freedom, that corresponds to $\chi^2$=25.0. The minimum $\chi^2$ are shown in fourth row.}\label{fig:specall.a-d.+LVG}
\end{figure}

\begin{figure}
  \renewcommand{\thefigure}{19}
  \begin{center}
    \FigureFile(77.9mm, 62.89mm){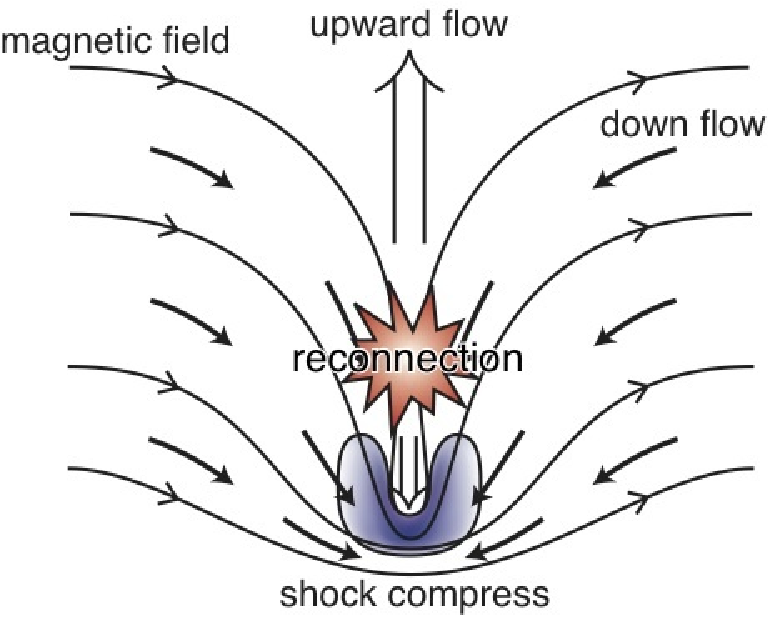}
  \end{center}
  \caption{Schematic image of reconnection at the foot point between two loops.}\label{fig:reconnection.sch}
\end{figure}


\begin{thebibliography}{}

\bibitem[Bania(1977)]{ban1977}	Bania,~T.~M. 1977, \apj, 216, 381
\bibitem[Bitran et al.(1997)]{bit1997}	Bitran,~M., Alvarez,~H., Bronfman,~L., May,~J., \& Thaddeus,~P. 1997, \aaps, 125, 99
\bibitem[Blake et al.(1987)]{bla1987}Blake,~G.~A., Sutton,~E.~C., Masson,~C.~R., \& Phillips,~T.~G. 1987, \apj, 315, 621
\bibitem[Dickman(1978)]{dic1978}	Dickman,~R.~L. 1978, \apjs, 37, 407
\bibitem[Caswell \& Haynes(1987)]{cas1987} Caswell,~J.~L., Haynes,~R.~F.\ 1987, \aap, 171, 261
\bibitem[Chieze, Pineau des Forets \& Flower (1998)]{chi1998} Chieze,~J.~P., Pineau des Forets,~G. \& Flower,~D.~R.\ 1998, Royal Astronomical Society, Monthly Notices, 295, 672C
\bibitem[Draine, Roberge, \& Dalgarno (1983)]{dra1983} Draine,~B.~T., Roberge,~W.~G. \& Dalgarno,~A.\ 1983, \apj, 264, 485
\bibitem[Ezawa et al. (2004)]{eza2004} Ezawa,~H., Kawabe,~R., Kohno,~K. and Yamamoto,~S. 2004, Proc. SPIE, 5489, 763
\bibitem[Ezawa et al. (2008)]{eza2008} Ezawa,~H., et al. 2008, Proc. SPIE, 7012, 701208
\bibitem[Frerking, Langer \& Wilson (1982)]{fre1982}Frerking,~M.~A., Langer,~W.~D. \& Wilson,~R.~W.\ 1982, \apj, 262, 590
\bibitem[Fujishita et al.(2009)]{fuj2009}Fujishita,~M. et al. 2009, \pasj, 61, 1039
\bibitem[Fukui et al.(1977)]{fuk1977}Fukui,~Y., Iguchi.~T., Kaifu,~N., Chikada,~Y., Morimoto,~M., Nagane,~K., Miyazawa,~W. \& Miyaji,~T. 1977, \pasj, 29, 643D
\bibitem[Fukui et al.(2006)]{fuk2006}	Fukui,~Y., et al.\ 2006, Science, 314, 106
\bibitem[Goldreich \& Kwan(1974)]{gol1974} Glodreich,~P. \& Kwan,~J. 1974, \apj, 189, 441
\bibitem[Goldsmith \& Langer (1978)] {gol1978} Glodsmith,~P.~F. \& Langer,~W.~D. 1978, \apj, 222, 881
\bibitem[G$\ddot{\mathrm{u}}$sten \& Henkel (1983)]{gus1983}G$\ddot{\mathrm{u}}$sten,~R., \& Henkel,~C. 1983, \aap, 136-145 
\bibitem[G$\ddot{\mathrm{u}}$sten \& Philipp(2004)]{gus2004}	G$\ddot{\mathrm{u}}$sten,~R., \& Philipp,~S.~D. 2004, in Proc.\ the Fourth Cologne-Bonn-Zermatt Symposium ed.\ S.~Pfalzner, C.~Kramer, C.~Staubmeier, A.~Heithausen (Heiderberg: Springer), 253
\bibitem[Handa et al. (1987)]{han1987}Handa,~T., Sofue,~Y., Nakai,~N., Hirabayashi,~H. \& Inoue,~M. 1987, /pasj, 39, 709
\bibitem[H\"uttemeister et al.(1993)]{hut1993}	H$\ddot{\mathrm{u}}$ttemeister,~S., Wilson,~T.~L., Bania,~T.~M. \& Mart\'in-Pintado,~J. 1993, \aap, 280, 255
\bibitem[Isobe, Tripathi \& Archontis (2007)]{iso2007} Isobe,~H., Tripathi,~D. \& Archontis,~V. 2007, \apj, 567L, 53
\bibitem[Kudoh \& Shibata(1999)]{kud1999} Kudoh,~T., \& Shibata,~K. 1999, \apj, 514, 493
\bibitem[Kurtz, Churchwell \& Wood (1994)]{kur1994}Kurtz,~S., Churchwell,~E. \& Wood,~D.~O.~S. 1994, \apjs, 91, 659
\bibitem[Kohno (2005)]{koh2005} Kohno,~K. 2005, ASP Conference series, 344, 242 
\bibitem[Langer \& Penzias (1990)]{lan1990} Langer,~W.~D. \& Penzias,~A.~A. 1990, \apj, 357, 477
\bibitem[Lis \& Goldsmith(1989)]{lis1989}	Lis,~D.~C., \& Goldsmith,~P.~F. 1989, \apj, 337, 704
\bibitem[Ladd et al.(2005)]{lad2005} Ladd,~N., Purcell,~C., Wong,~T. \& Robertson,~S.\ 2005, PASA, 22, 62
\bibitem[Leung, Herbst \& Huebner(1984)]{leu1984}Leung,~C.~M., Herbst,~E., \& Huebner,~W.~F.\ 1984, 56, 231
\bibitem[Lockman(1989)]{loc1989} Lockman,~F.~J.\ 1989, \apjs, 71, 469L
\bibitem[Machida et al.(2009)]{mac2009}	Machida,~M. et al. 2009, \pasj, 61, 441
\bibitem[Martin et al.(2004)]{mar2004}	Martin,~C.~L., Walsh,~W.~M., Xiao,~K., Lane,~A.~P., Walker,~C.~K. \& Stark,~A.~A. 2004, \apjs, 150, 239 
\bibitem[Matsumoto et al.(1988)]{mat1988}	Matsumoto,~R., Horiuchi,~T., Shibata,~K. \& Hanawa,~T. 1988, \pasj, 40, 171
\bibitem[Mezger \& Henderson (1967)]{mez1967} Mezger,~P.~G. \& Henderson,~A.~P. 1967, \apj, 147, 471 
\bibitem[Miyamoto \& Nagai(1975)]{miy1975}	Miyamoto,~M., \& Nagai,~R. 1975, \pasj, 27, 533
\bibitem[Morris \& Serabyn(1996)]{mor1996}	Morris,~M., \& Serabyn,~E. 1996,  \araa, 34, 645
\bibitem[Nagai et al.(2007)]{nag2007} Nagai,~M., Tanaka.~K., Kamegai,~K. \& Oka,~T.\ 2007, \pasj, 59, 25
\bibitem[Oka et al.(2005)]{oka2005}	Oka,~T., Geballe,~T.~R., Goto,~M., Usuda,~T. \& McCall,~B.~J. 2005, \apj, 632, 882
\bibitem[Oka et al. (2007)]{oka2007} Oka,~T., Nagai,~M., Kamegai,~K., Tanaka,~K. \& Kuboi,~N. 2007, \pasj, 59, 15
\bibitem[Oka et al.(1998)]{oka1998}	Oka,~T., Hasegawa,~T., Hayashi,~M., Handa,~T., \& Sakamoto,~S. 1998, \apj, 493, 730
\bibitem[Panagia (1973)]{pan1973} Panagia,~N. 1973, \aj, 78, 929
\bibitem[Parker(1966)]{par1966}Parker,~E.~N. 1966, \apj, 145, 811
\bibitem[Riquelme et al.(2010)]{riq2010} Riquelme,~D., Amo-Baladron,~M.~A., Martin-Pintado,~J., Mauersberger,~R., Bronfman,~L. \& Martin,~S. (2010) Proceedings of the Galactic Center Workshop 2009, Shanghai.
\bibitem[Rodr$\acute{\i}$guez-Fern$\acute{\mathrm{a}}$ndez et al.(2001)]{rod2001}	Rodr$\acute{\i}$guez-Fern$\acute{\mathrm{a}}$ndez,~N.~J., Mart$\acute{\i}$n-Pintado,~J., Fuente,~A., de Vicente,~P., Wilson,~T.~L. \& H$\ddot{\mathrm{u}}$ttemeister,~S. 2001, \aap, 365, 174
\bibitem[Scoville \& Solomon (1974)]{sco1974} Scoville,~N,~Z., \& Solomon,~P.~M. 1974, \apj, 187, L67
\bibitem[Scoville, Solomon \& Penzias (1975)]{sco1975} Scoville,~N.~Z., Solomon,~P.~M., \& Penzias,~A.~A. 1975, \apj, 201, 352
\bibitem[Sofue(1996)]{sof1996}	Sofue,~Y. 1996, \apj, 458, 120
\bibitem[Takahashi et al.(2009)]{tak2009}Takahashi,~K. et al. 2009, \pasj, 61, 957
\bibitem[Torii et al. (2009)]{tor2009}Torii,~K. et al. 2009, submitted to \pasj
\bibitem[Tsuboi, Handa \& Ukita (1999)]{tsu1999} Tsuboi,~M., Handa,~T. \& Ukita,~N. 1999, \apjs, 120, 1
\bibitem[Wang et al.(1994)]{wan1994} Wang,~Y., Jaffe,~D.~T., Graf,~U.~U. \& Evans,~N.~J.,II\ 1994, \apjs, 95, 503
\bibitem[Wilson \& Rood (1994)]{wil1994} Wilson,~T.~L. \& Rood,~R. 1994, \araa, 32, 191

\end{thebibliography}
\end{document}